\documentclass[onecolumn]{emulateapj}
\slugcomment{Astro-ph archives (submitted: 10$^{th}$ April 2015)}
\shorttitle{Contraction and disruption of coronal loops and associated M6.2 flare}
\shortauthors{Kushwaha et al.}
\begin{document}
\title{Large-scale contraction and subsequent disruption of coronal loops\\ during various phases of the M6.2 flare associated with\\ the confined flux rope eruption}
\author{Upendra Kushwaha\altaffilmark{1}}

\affil{Udaipur Solar Observatory, Physical Research Laboratory, Udaipur 313001, India}
\email{upendra@prl.res.in}
\altaffiltext{1}{Indian Institute of Technology Gandhinagar, Ahmedabad 382424, Gujarat, India}
\author{Bhuwan Joshi\altaffilmark{2}}
\affil{School of Space Research, Kyung Hee University, Yongin, Gyeonggi-Do, 446-701, Korea}
\altaffiltext{2}{On sabbatical leave from Udaipur Solar Observatory, Physical Research Laboratory, Udaipur 313001, India}

\author{Astrid M. Veronig}
\affil{Kanzelh\"ohe Observatory/Institute of Physics, University of Graz, Universit$\ddot{a}$tsplatz 5, A-8010 Graz, Austria}

\author{Yong-Jae Moon}
\affil{School of Space Research, Kyung Hee University, Yongin, Gyeonggi-Do, 446-701, Korea}

%
%

\begin{abstract}

We present a detailed multi-wavelength study of the M6.2 flare which was associated with a confined eruption of a prominence using TRACE, RHESSI, and NoRH observations. The pre-flare phase of this event is characterized by spectacular large-scale contraction of overlying extreme ultraviolet (EUV) coronal loops during which the loop system was subjected to an altitude decrease of $\sim$20~Mm (40\% of the initial height) for an extended span of $\sim$30 min. This contraction phase is accompanied by sequential EUV brightenings associated with hard X-ray (HXR) (up to 25 keV) and microwave (MW) sources from low-lying loops in the core of the flaring region which together with X-ray spectra indicate strong localized heating in the source region before the filament activation and associated M-class flare. With the onset of the impulsive phase of the M6.2 flare, we detect HXR and MW sources that exhibit intricate temporal and spatial evolution in relation with the fast rise of the prominence. Following the flare maximum, the filament eruption slowed down and subsequently confined within the large overlying active region loops; the event did not lead to a coronal mass ejection (CME). During the confinement process of the erupting prominence, we detect MW emission from the extended coronal region with multiple emission centroids which likely represent emission from hot blobs of plasma formed after the collapse of the expanding flux rope and entailing prominence material. RHESSI observations reveal high plasma temperature ($ \sim $30 MK) and substantial non-thermal characteristics with electron spectral index ($\delta \sim $5) during the impulsive phase of the flare. The time-evolution of thermal energy exhibits a good correspondence with the variations in cumulative non-thermal energy which suggest that the energy of accelerated particles efficiently converted to hot flare plasma implying an effective validation of the Neupert effect. 

\end{abstract}


\section{Introduction}

The eruption of prominences and magnetic flux ropes from the Sun has important implications in view of space weather manifestations. These eruptions are frequently associated with solar flares and coronal mass ejections. It is widely accepted that magnetic reconnection is the most fundamental process responsible for the changes in the topology of magnetic fields, as well as the rapid conversion of stored magnetic energy into thermal and kinetic energy of plasma and particles during solar eruptive events \citep[for a review, see][]{Priest2002}.

Solar flares occur in active regions consisting of sunspot groups in the photosphere enveloped by large overlying coronal loops that essentially represent a closed magnetic field configuration associated with sunspots. The spectrum of energy released from a flare shows that there is a precursor phase, which is generally marked by a gradual increase in thermal radiation mainly in the form of small enhancement of soft X-ray (SXR) flux \citep{Veronig2002}. It is followed by the impulsive phase lasting for a few seconds to tens of minutes, comprising of energy release in the form of $\gamma$-ray, hard X-ray (HXR), extreme ultraviolet (EUV), and microwave (MW) emissions. After the impulsive phase thermal radiation dominates. In this phase, H$\alpha$ and SXR emission attain a maximum level followed by a gradual decline in the intensity. To understand the diversity of phenomena during eruptive flares, the standard flare model (also known as CSHKP model) has been proposed which incorporates the pioneering works of \cite{Charmichael1964}, \cite{Sturrock1966}, \cite{Hirayama1974}, and \cite{Kopp1976}. 

In the standard flare model, the initial driver of the flare process is a rising prominence (also called filament when viewed on the solar disk) formed in the chromosphere along the magnetic polarity inversion line of solar active regions. In a simplistic picture, one can imagine the structure of a bipolar magnetic configuration in terms of an inner region, called the core field, and the outer region, called the envelope field. The core fields are rooted close to the neutral line while the envelope fields are rooted away from it. Many observations have confirmed that core fields can usually be traced by a filament. The standard flare model recognizes that in the early stages of a large eruptive flare, the core fields containing the prominence erupt which would result in the stretching of the envelope fields. With the evolution of the eruption process, the stretched field lines reclose via magnetic reconnection beneath the erupting filament. The extensive solar observations over the last decades, mainly from space borne telescopes, have revealed several important features of the multi-wavelength flare that have further strengthened the standard flare model and provided concrete observational inputs to understand the role of magnetic reconnection toward plasma heating and particle acceleration. These features include: occurrence of hard X-ray (HXR) footpoint (FP) and looptop (LT) sources, formation of H$\alpha$ flare ribbons at the opposite polarity regions, rising arcade of the intense soft X-ray (SXR) loops, plasmoid ejections, upward motion of HXR LT source, and increasing separation of H$\alpha$ flare ribbons along with HXR FP sources \citep[see reviews by][]{Moore1992,Shibata1998,Benz2008,Fletcher2011,Karlicky2014}.

According to the standard flare model, the filament activation is a crucial part of the flare process. Observations reveal several categories of the filament activation, viz. full, partial and failed \citep{Gilbert2007}. The full or partial eruptions are followed by flares and CMEs while the failed eruptions characterize those situations when a filament undergoes a successful activation in the beginning but finally fails to erupt from the corona. Full and partial eruption of filaments have been reported in numerous observations. On the other hand, failed eruptions are observed less frequently and therefore form a less studied topic in solar physics. Their study provide an opportunity to investigate the interaction between the erupted plasma and magnetic fields with the overlying coronal loops \citep[see e.g.,][]{Rust2003,Mrozek2011,Kumar2011,Netzel2012}. The rate of decrease of overlying magnetic fields with their heights is considered to be a crucial factor that determines whether an erupting prominence will lead to a successful eruption or confine within the overlying coronal region \citep{Torok2005}. \cite{Ji2003} and \cite{Alexander2006} studied a failed eruption of a filament during which multiple HXR sources were observed from the coronal as well as footpoint locations. The examination of the locations of coronal HXR sources with respect to the dynamical evolution of the filament is of high interest toward a better understanding of the energy release scenario in solar eruptions.    

Sometimes prior to the filament activation and associated flares, small localized brightenings are observed to occur in the active region close to the site of the main flare \citep{Kundu2004,Chifor2007,Joshi2013,Zhang2014}. Such localized pre-flare brightenings are frequently observed as enhanced X-ray flux at low energies (e.g., GOES light curves) before the impulsive phase. It has been recognized that filaments undergo slow yet important changes during the pre-flare phase \citep{Chifor2006,Chifor2007,Joshi2011,Joshi2013,Awasthi2014}. Some studies also show crucial evidence of non-thermal X-ray emission during the early activation phase of solar eruptions \citep{Farnik2003,Joshi2011}. These investigations have provided clues to understand the processes occurring in the active region that eventually lead to filament destabilization and thus triggering the impulsive energy release. It is important to note that, while investigating the pre-flare phase, one should also carefully probe the state of overlying coronal loops besides the core region containing the filament.   

In this study, we present observations of an M6.2 flare and its associated pre-flare activities observed in active region NOAA 10646 on 2004 July 14 between 04:30 UT and 06:00 UT that are associated with the confined eruption of a filament (SOL2004-07-14). During the pre-flare phase, we observe contraction of overlying coronal loops as well as multiple brightenings at its core region where the main flare occurred. The impulsive phase of the M6.2 flare is characterized by multiple non-thermal peaks during which a rising prominence showed an initial acceleration for $ \sim $4 minutes, then slowed down and finally stopped by overlying loops. Several aspects of this event related to the confinement of the prominence eruption and associated flare emissions have been analyzed in previous studies \citep{Mrozek2011,Netzel2012}. \citet{Mrozek2011} reported that this flare can be well described by a quadrupole model. This study showed that the failed eruption of the flux rope caused stretching in the overlying loop system. As a consequence of this interaction, the overlying loops were observed to oscillate radially with an initial amplitude of 9520 km, a period of 377 s and an exponential damping time of 500 s. This flare is one of three events analyzed in \citet{Netzel2012} in which they found a close relation among the EUV brightenings observed at the footpoints of overlying loop systems, a decrease in velocity of the eruption, and episodes of hardening of the HXR spectra. From these observations, they concluded that the interaction of the erupting structure with overlying loops was strong enough to heat the high loops which then cooled down in $ \sim $30-60 minute to become visible in the EUV range.

Here we revisit SOL2004-07-14 and provide a comprehensive analysis of the intriguing flaring and eruptive activities from its prolonged pre-flare phase to the decay of the event. We obtained significant additional information about the state of overlying coronal loops, the flaring core region and energy release processes during the flare's impulsive phase. On the whole, this event was remarkable in several aspects. (1) We note significant pre-flare activity which is manifested in the form of large-scale contraction of overlying coronal loops. (2) Accompanying with the large-scale loop contraction, sequential brightening in the core region was observed which can clearly be recognized as three pre-flare events in the GOES X-ray profile. (3) A drastic increase in the speed of contraction is observed just $\sim $5 minute before the onset of the impulsive phase (i.e., during third pre-flare event). (4) During the impulsive phase, we observe multiple non-thermal peaks. The first stage of the impulsive phase marks the transition from signs of the magnetic implosion scenario \citep{Hudson2000} to the generally observed expansion of flaring coronal loops and the activation of filament. The high spatial resolution, continuous sequence of EUV images at 171~\AA~from the pre-flare phase to the end of the flare provided us with an excellent opportunity to thoroughly examine activities in the core region of the eruption along with the overlying system of coronal loops. The signatures of magnetic reconnection and particle acceleration are inferred from HXR and microwave (MW) observations. We further provide a detailed HXR spectroscopy of the flare to investigate the crucial plasma parameters and flare energetics. The paper is organized as follows. In Section~\ref{data}, we provide details of observing instruments and data analysis procedures. For the convenience of readers, an overview of the multi-wavelength observations are presented in Section~\ref{overview}. We provide a detailed multi-wavelength diagnostic of each phase of the eruption in Section~\ref{results}. In Section~\ref{discuss}, we discuss and interpret our observations. The summary of the paper is given in the final section.     

\begin{figure}
\epsscale{0.85}
\plotone{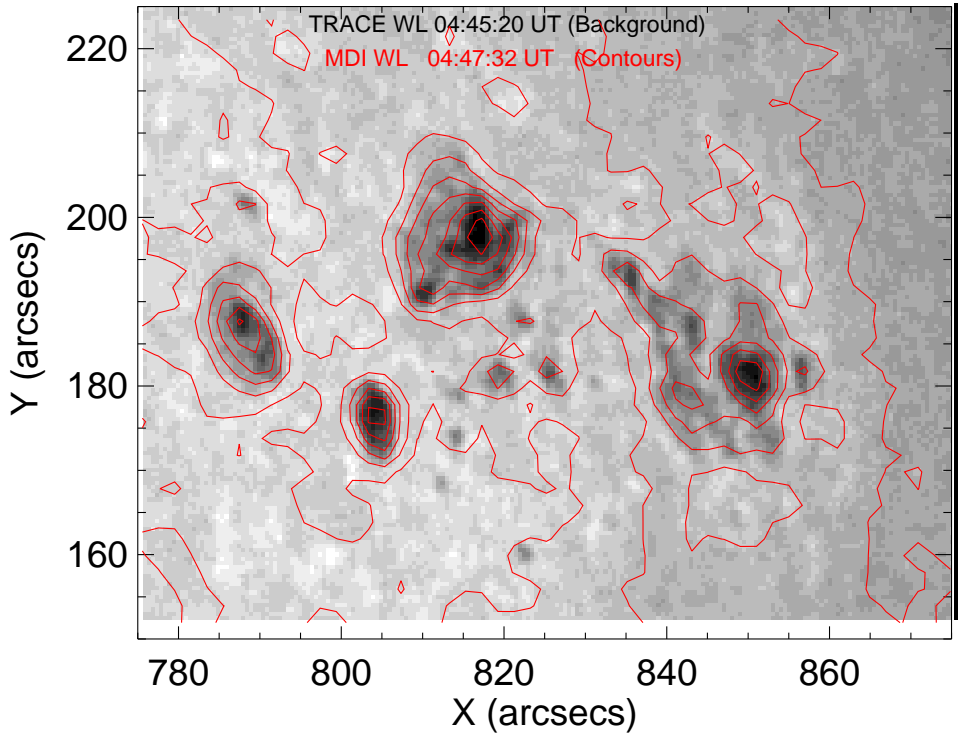}
\caption{Illustration of the alignment of nearly co-temporal TRACE and SOHO/MDI white light images.}
\label{pointing}
\end{figure} 
\bigskip
\section{Data}
\label{data}

In this paper, we present a comprehensive multi-wavelength analysis of the sequence of intriguing activities that occurred in solar active region NOAA 10646 on 2004 July 14 between 04:30 UT and 06:00 UT. Transition Region and Coronal Explorer (TRACE; \citealt{Handy1999}) observed the active region NOAA 10646 during this whole period at high time cadence (better than 1 min) with its EUV channel at 171 \AA. The TRACE telescope has a field of view of 8$ \arcmin $.5$ \times $8$ \arcmin $.5 and a spatial resolution of 1$ \arcsec $ (0$ \arcsec $.5 pixel$ ^{-1} $). The TRACE 171 \AA~filter is mainly sensitive to plasmas at a temperature around 1 MK (Fe IX/X).

Reuven Ramaty High Energy Solar Spectroscopic Imager (RHESSI; \citealt{Lin2002}) detected significant activities in NOAA 10646 during its continuous observations from 04:30 UT to 05:35 UT. RHESSI observes the full Sun with an unprecedented combination of spatial resolution (as fine as $ \sim $2$ {\arcsec} $.3) and energy resolution (1--5 keV) in the energy range of 3 keV to 17 MeV. To reconstruct X-ray images, we have used the computationally expensive PIXON algorithm \citep{Metcalf1996,Hurford2002} which is thought to provide the most accurate image photometry \citep{Alexander1997}. The images are reconstructed by selecting front detector segments 3--8 (excluding 7) with 20 s integration time.

Our analysis involves the identification of X-ray and MW source locations during the dynamical evolution of the filament and coronal loops observed by TRACE. However, the pointing information of TRACE is often not good enough to obtain a correct co-alignment between TRACE images and imaging data at other wavelengths. For a full disk imager (such as SOHO/MDI; \citealt{Scherrer1995}), the solar limb provides a reference that helps in accurate determination of the absolute pointing of the telescope. Moreover, thermal bending of the TRACE guide telescope also leads to an unknown variation in the pointing of at least a few arcseconds \citep{Fletcher2001, Alexander2006}. The pointing information of both RHESSI and SOHO is believed to be sufficiently accurate. Therefore, we corrected TRACE pointing by a cross-correlation alignment between a TRACE white light (WL) and a SOHO/MDI WL image taken at 04:45:20 UT and 04:47:32 UT respectively using the method described in \citet{Gallagher2002}\footnote{https://www.tcd.ie/Physics/people/Peter.Gallagher/trace-align/index.html}. This method has been found very effective for near-limb events \citep[see, e.g.,][]{Liu2009,Joshi2013}. We found that the TRACE pointing was offset by 0$ \arcsec $.8 in the X direction and 2$ \arcsec $.4  in the Y direction (see Figure \ref{pointing}).

In this analysis, we include microwave (MW) observations by the Nobeyama Radioheliograph (NoRH; \citealt{Nakajima1994}) at 17 GHz and 34 GHz. At these two frequencies NoRH has spatial resolutions of 10$ \arcsec $ and 5$ \arcsec $ respectively \citep{Takano1997}. NoRH has a sensitivity of at least 1 solar flux unit (sfu) at 17 GHz and $ \sim $3 sfu at 34 GHz. The normal time resolution of NoRH is 1 s, but a resolution as good as 50 ms can be used for special projects \citep[see also][]{Kundu2006}. For the present study we have analyzed NoRH light curves and images with a temporal cadence of 1 s.
                   
\section{Event overview: phases of flare evolution}
\label{overview}

\begin{figure}
\vspace*{-1.0cm}
\epsscale{0.9}
\plotone{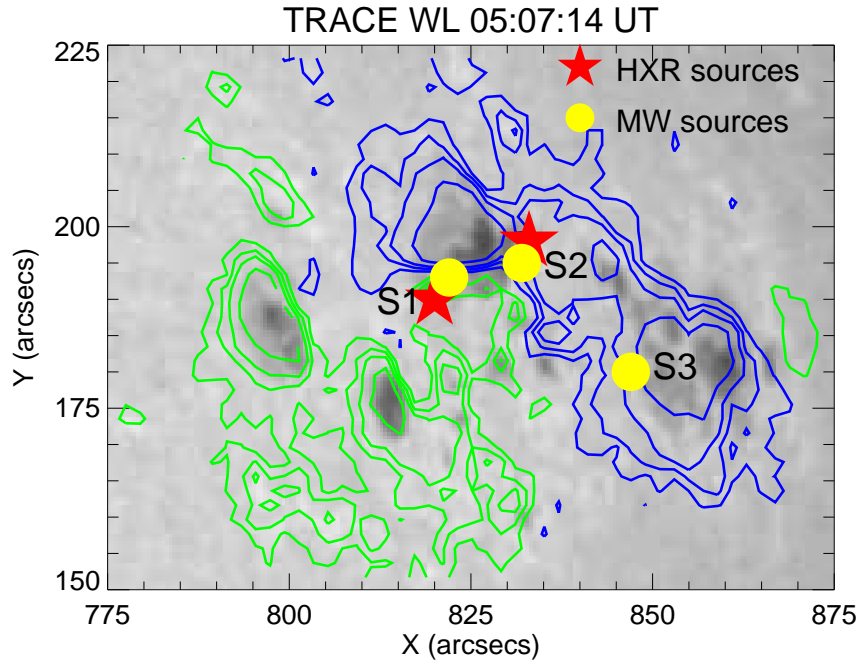}
\vspace*{-1.0cm}
\caption{TRACE white light (WL) image overlaid by contours of co-temporal magnetogram taken from SOHO/MDI. The contour levels are $ \pm $100, $ \pm $200, $ \pm $300, $ \pm $500, $ \pm $800, $ \pm $1000 and $ \pm $1500 Gaus for positive polarity magnetic flux (blue) and negative polarity magnetic flux (green) respectively. The location of HXR and MW sources during the flare are indicated by star ($\bigstar$) and circle ($ \bullet $) respectively. We mark two conjugate emission centroids of HXR and MW sources observed during early impulsive phase of the M6.2 flare (see also Figures \ref{MainEvent}(b) and \ref{tr_norh}(b)) with S1 and S2 respectively. We further indicate a distant MW source by S3 (see also Figure \ref{tr_norh}(d)) which originated during the main impulsive phase (i.e., at 05:21 UT) and prevailed till the end of flare ($ \sim $05:30 UT). MW source S3 does not have any HXR counterpart.}   
\label{mdi_hsi}
\end{figure}

On 2004 July 14, active region NOAA 10646 was situated close to the western limb at heliographic coordinates of N13W64. In Figure \ref{mdi_hsi}, we show a TRACE white light (WL) image overlaid by a SOHO/MDI line-of-sight (LOS) magnetic field map of NOAA 10646. We find that the active region consists of a set of sunspot groups with leading and trailing groups having positive and negative magnetic polarities respectively. The active region displayed a rather dispersed structure of sunspots with well separated magnetic polarities.

\bigskip

\begin{figure}
\epsscale{1.0}
\plotone{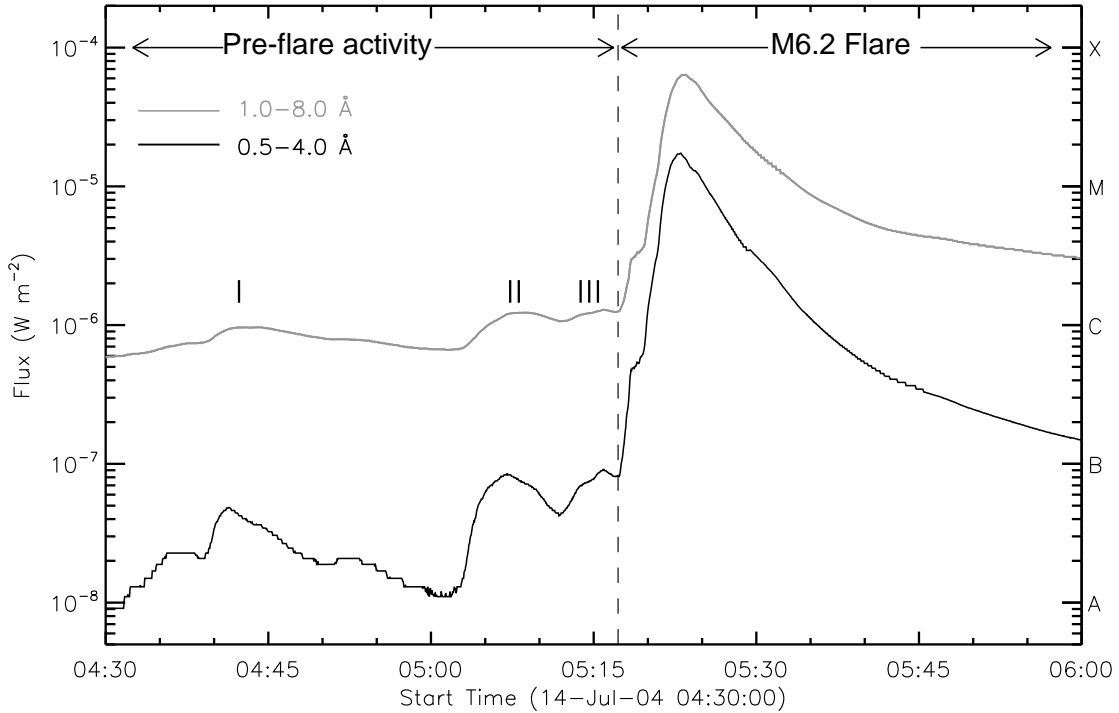}
\caption{GOES soft X-ray flux profiles in 0.5$-$4 \AA~and 1$-$8~\AA~channels between 04:30 and 06:00~UT. The vertical--dashed line distinguish the pre-flare activity from the main M6.2 flare. The pre-flare phase (04:30 UT -- 05:17 UT) is characterized by multiple small peaks, which are more pronounced in GOES high energy channel (i.e., in 0.5--4.0 \AA) and marked by I, II, and III.}   
\label{goes_profile}
\end{figure}

The active region NOAA 10646 exhibited a set of very intriguing flaring and eruptive activities between 04:30 UT and 06:00 UT which can be nicely followed in GOES SXR time profiles shown in Figure~\ref{goes_profile}. A major flare of class M6.2 occurred in the active region between 05:17 and 06:00 UT. It is remarkable to observe significant activities in the active region that started several minute before the onset of the M-class flare that continued until the flare's impulsive phase. The GOES light curves (Figure \ref{goes_profile}) clearly recognizes these pre-flare activities in the form of multiple small peaks at $ \sim $04:41 UT, $ \sim $05:06 UT and $ \sim $05:16 UT before the onset of the impulsive phase of flare at $ \sim $05:17 UT. Thereafter, the flare attains its peak at $ \sim $05:23 UT and gradually ended by 06:00 UT. We mark pre-flare GOES peaks as I, II, and III in Figure \ref{goes_profile}. These activities correspond to the localized and discrete brightenings at low-lying loops at the site of the main flare and simultaneous contraction of large overlying coronal loops. The detailed imaging analysis of X-ray and EUV data presented in the next section clearly show that the pre-flare activity corresponds to phases of episodic energy release at the earliest stages of restructuring of magnetic fields. It is further noteworthy that multiple peaks at pre-flare phases are more pronounced in the high energy GOES channel (i.e., in 0.5--4.0 \AA~band). Based on GOES and TRACE observations, we consider following phases of flare evolution: (1) pre-flare activity which includes large-scale loop contraction and brightenings in the core region (2) M6.2 flare which is accompanied with a failed eruption of a filament. Here we note that although the loop contraction proceeds during the whole pre-flare activity, we describe loop contraction and core region brightenings separately in the next section to provide a better clarity on these two phenomena. We should note that the contraction was observed in higher overlying coronal loops that are spatially separated from the regions of pre-flare brightenings. It is, however, quite possible that the two phenomena (localized brightenings and contraction of overlying loop system) are physically interconnected.      

In Figure \ref{mdi_hsi}, we further show the locations of a pair of HXR and MW emission centroids observed during the early impulsive phase of M6.2 flare by star ($\bigstar$) and filled circle ($ \bullet $) symbols respectively. It is apparent that the flare occurred close to the neutral line of photospheric magnetic field. We mark locations of HXR and MW emission centroids as S1 and S2. Following the main impulsive phase, we note another prominent MW source appeared at a distant location, marked as S3 in Figure \ref{mdi_hsi}, which does not have any HXR counterpart.         
 
\section{Analysis and results}
\label{results}
\subsection{Pre-flare activity}
\subsubsection{Large-scale contraction of coronal loops}
\label{sec:cont}
\begin{figure}
\vspace*{-1.0cm}
\epsscale{0.9}
\plotone{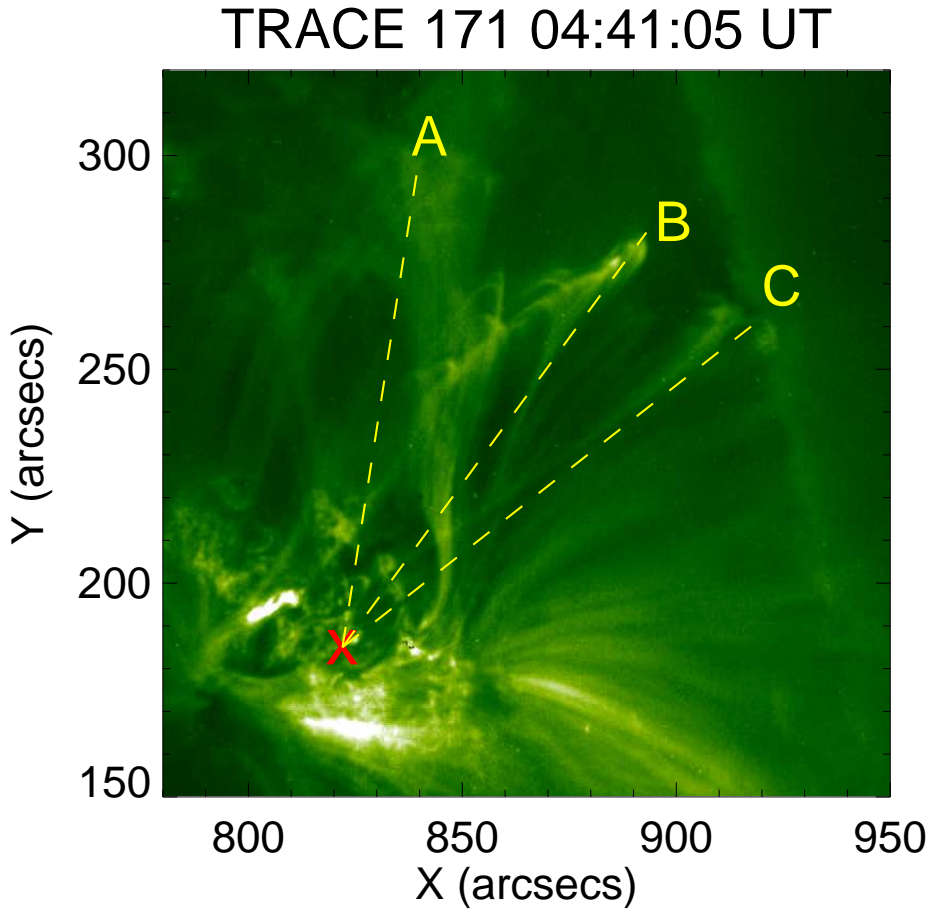}
\vspace*{-0.5cm}
\caption{TRACE 171 \AA~image showing a set of coronal loops associated with the active region NOAA 10646. Three system of distinct sets of loops can be recognized and are named as A, B, and C. Their heights from the source region are estimated as $ \sim $92 Mm, 88 Mm, and 93 Mm. Only the loop system `B' exhibited a large-scale contraction between 04:47 and 05:18 UT.}
\label{loops}
\end{figure}

\begin{figure}
\epsscale{1.05}
\plotone{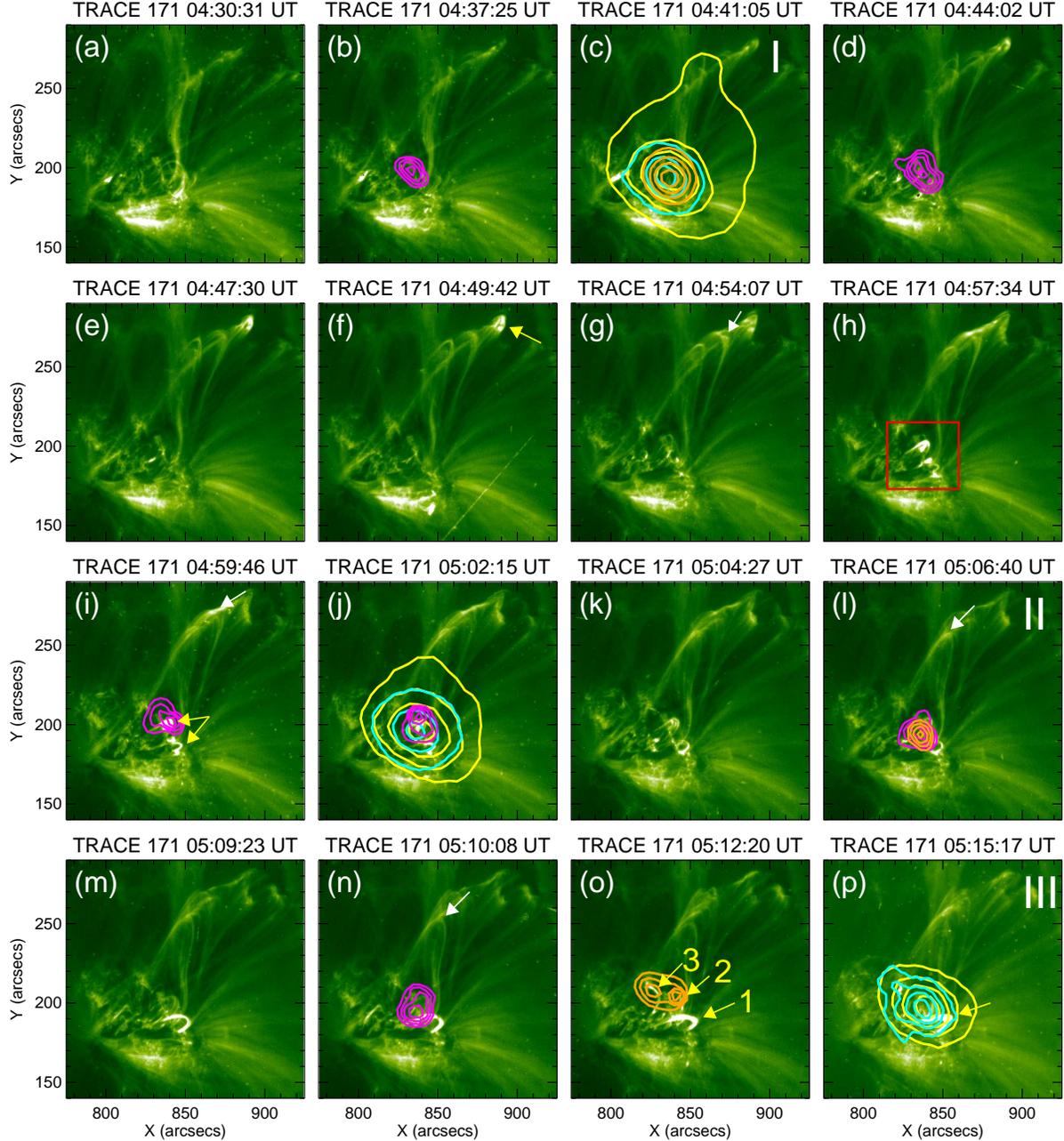}
\caption{Series of TRACE 171~\AA~images during the pre-flare phase of the M6.2 flare showing the contraction of coronal loops as well localized brightening at the source region. We have marked I, II and III in panels (c), (l), and (p) respectively to show EUV brightenings at the events of GOES flux enhancement (see Figure \ref{goes_profile}) before the M6.2 flare onset. We indicated the contracting loops by the white arrows in a few panels. An intense brightening is noted in the high corona and indicated in panel (f). The sequential brightenings in the low-lying coronal loops are enclosed within a box in panel (h) and their evolution during later stages are indicated by arrows in panels (i) and (o). Co-temporal X-ray (6$-$12 keV: magenta, 12$-$25 keV: orange) and MW (17 GHz: yellow, 34 GHz: sky) contours are over plotted on a few selected panels. The contour levels are set as 15, 30, 50, and 80\% of the peak flux for both X-ray and MW images.}
\label{pre-2}
\end{figure}

\begin{figure}
\epsscale{1.00}
\plotone{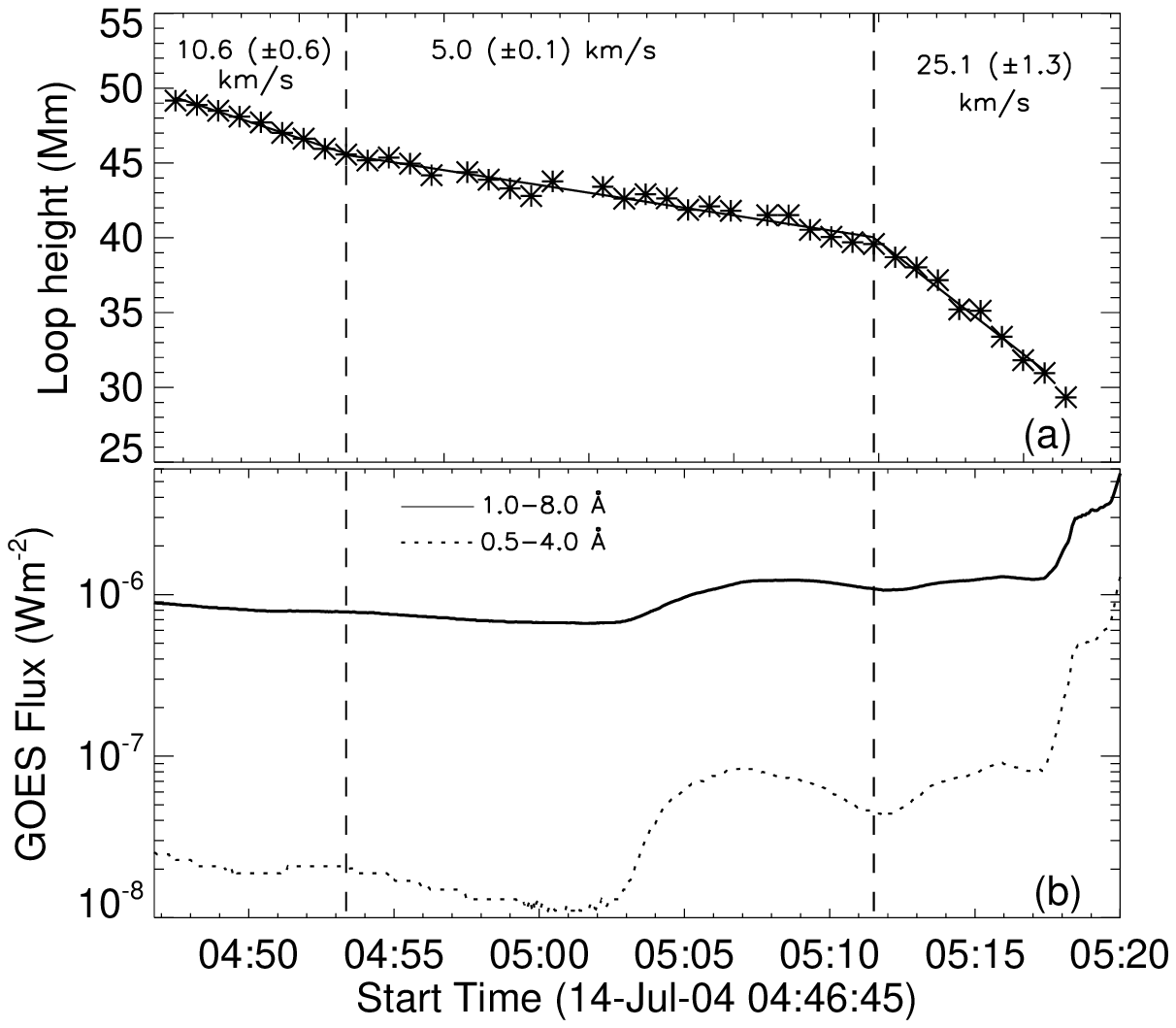}
\caption{Height-time plot of the coronal loop system presenting contraction motion between 04:47 UT and 05:18 UT (panel (a)). The speed of downward motion of the contracting loops during different phases are also indicated with their respective uncertainty in the measurements. To compare the evolutionary phases of loop contraction with the flare evolution, we plotted the GOES flux profile for the same time interval in the panel (b).}
\label{imp_ht}
\end{figure}

Active region NOAA 10646 was enveloped by a set of large coronal loops with different orientations which present a peculiar structure of active region corona (see Figure \ref{loops}). We have estimated the height of three prominent overlying loop systems (which are indicated in Figure \ref{loops} by A, B, and C) from the source region (shown by `$ \times $'). The projected height of these three systems of loops were estimated as $ \sim $92 Mm, 88 Mm, and 93 Mm. We found significant activities occurring in the coronal as well as source region by examining 171 \AA~images available from 04:30 UT onward. From the sequence of EUV images, it is evident that the higher coronal loops began to collapse from $ \sim $04:47 UT (see Figures \ref{pre-2} \& \ref{imp_ht}; and online movie). We find that only one loop system `B' underwent contraction. This large-scale contraction continued until the impulsive rise of the M6.2 flare at 05:18 UT.

\bigskip

We note some important observational characteristics of the loop contraction. (1) Before and at the beginning of loop contraction (i.e., 04:42--04:51 UT) the top of the loop system exhibited intense emission which gives evidence for localized heating in the higher corona (indicated by the arrow in Figure \ref{pre-2}(f)). (2) From 04:51 UT onward, this loop top brightening started getting dispersed and the brightening mostly dissolves by $ \sim $05:02 UT. (3) From $ \sim $4:54 UT, we observe intermittent brightenings in a set of low-lying loops at the source region besides contraction of higher loops (Figure \ref{pre-2}(h)).

In Figure \ref{imp_ht} (a), we present a height-time plot of the collapsing coronal loops. Due to mingling of several loops which altogether evolve dynamically, the estimation of loop height was a tedious exercise. We took utmost care in distinguishing one of the loop systems undergoing the contraction by carefully identifying its legs. We find that the loop contraction proceeds with varying speed. In order to compare the phases of loop contraction with the thermal emission from the flaring region, we have also plotted the co-temporal $GOES$ flux profiles in both channels in Figure \ref{imp_ht}(b). We find that in terms of contraction speed, the collapsing loop system undergoes three phases of evolution. In the beginning, the loops contract with relatively higher speed of $\sim $11~km~s$^{-1}$ that sustained for $\sim$6 minutes. This is followed by a prolonged phase of $ \sim $19 minutes during which the contraction slowed down to $ \sim $5 km s$ ^{-1} $. As shown in Figure \ref{goes_profile}, the pre-flare activity is characterized by three episodes of flux enhancement (marked as I, II, and III). We note that the loop contraction continued during all three stages (Figure \ref{imp_ht}) but there is a drastic increase in the speed of contraction from $ \sim $5 km s$ ^{-1} $ to $\sim$25~km s$^{-1}$ at $\sim$5:12 UT which mark the third episode of flux enhancement. The loop contraction halts at 05:18 UT with the beginning of the flare impulsive phase after which the expansion predominates over the implosion. 

\subsubsection{Localized brightenings at core region}
\label{pre-flare}

The sequence of images during the pre-flare phase clearly indicate that the large-scale contraction of EUV coronal loops is accompanied by localized brightenings at the core of active region loops. In Figure \ref{pre-2}, we show EUV images showing pre-flare activity and mark the three events of GOES flux enhancement (see Figure \ref{goes_profile}) as I, II and III in panels (c), (l), and (p) respectively. Here it is important to note that the core region exhibited pre-flare emission even before the onset of contraction in the form of X-ray (6--12 keV and 12--25 keV) and MW (17 GHz and 34 GHz) emissions which are co-spatial (Figures~\ref{pre-2}(b)-(d)). However, at this time EUV images do not show any significant bright structures associated with X-ray and MW emitting regions. The X-ray and MW sources in the pre-flare phase clearly indicate distinct events of energy release at the core of active region loops that precede the large-scale contraction of overlying loops.

We observe localized, distinct bright EUV sources at the core region from 04:54 UT onward, i.e., after the initiation of the contraction of the loop system. The location of EUV brightening is shown inside a box in Figure~\ref{pre-2}(h). Although brightening at three locations were observed in the beginning ($ \sim $04:57 UT), we can clearly recognize two of the three bright structures as low-lying loops at some later stages ($ \sim $05:02 UT). The intensity of bright loops rapidly decreased ($ \sim $05:07 UT) and we observe formation of a relatively large system of low-lying loops thereafter (i.e., 05:08 UT onward). In fact three new bright loops formed sequentially which appeared in coincidence for a brief interval (around 05:12 UT). These three loops are marked as 1, 2, and 3 in order of their occurrences. We also note that the loop--1 became very intense in the later stages (marked in Figure \ref{pre-2}(p) at 05:15:17 UT) and also lasted till the impulsive phase of M6.2 flare. The comparison of pre-flare EUV brightening with the co-temporal RHESSI images clearly reveal X-ray sources at the top of low-lying loops. It is interesting to see that although in EUV images, loop--1 appeared the most intense, the HXR emission in 12--25 keV energy band is essentially associated with loops 2 and 3. We further note that MW emission at 17 GHz and 34 GHz is observed in the form of single source that shows clear association with the bright low-lying EUV loops (Figure \ref{pre-2}(p)).

\subsection{The M6.2 flare and failed eruption}
\subsubsection{Multi-wavelength observations}

\begin{figure}
\epsscale{0.9}
\plotone{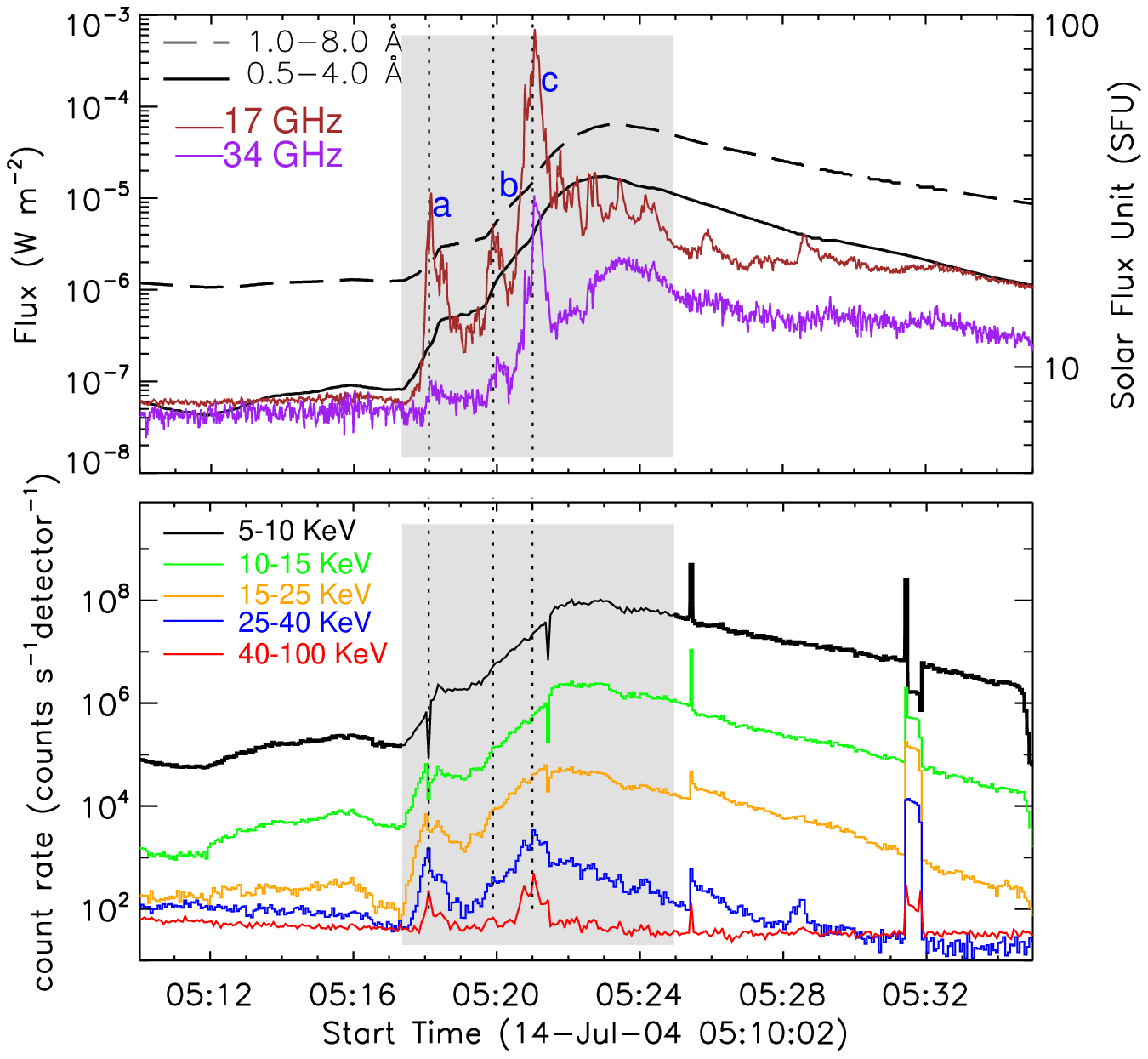}
\caption{Top panel: GOES soft X-ray flux profiles in the 0.5--4 \AA~and 1--8 \AA~chanels between 05:10 UT and 05:36 UT. The figure also presents time evolution of microwave emission at 17 GHz and 34 GHz frequencies, obtained from NoRH. Bottom panel: The temporal variations of RHESSI X-ray fluxes in five different energy bands. For the clarity in presentation, we have scaled RHESSI count rates by factor of 1, 1, 1/2, 1/5, and 1/40 for 5--10, 10--15, 15--25, 25--40, and 40--100 keV energy bands respectively. Note that the impulsive phase of the the flare is indicated by grey shaded area during which MW emission exhibits three events of flux enhancement `a', `b', and `c'. The events `a' and `c' are common in HXR and MW profiles (above 25 keV) while event `b' does not appear prominent in HXR observations.}
\label{flux_profile}
\end{figure}

\begin{figure}
\epsscale{0.95}
\plotone{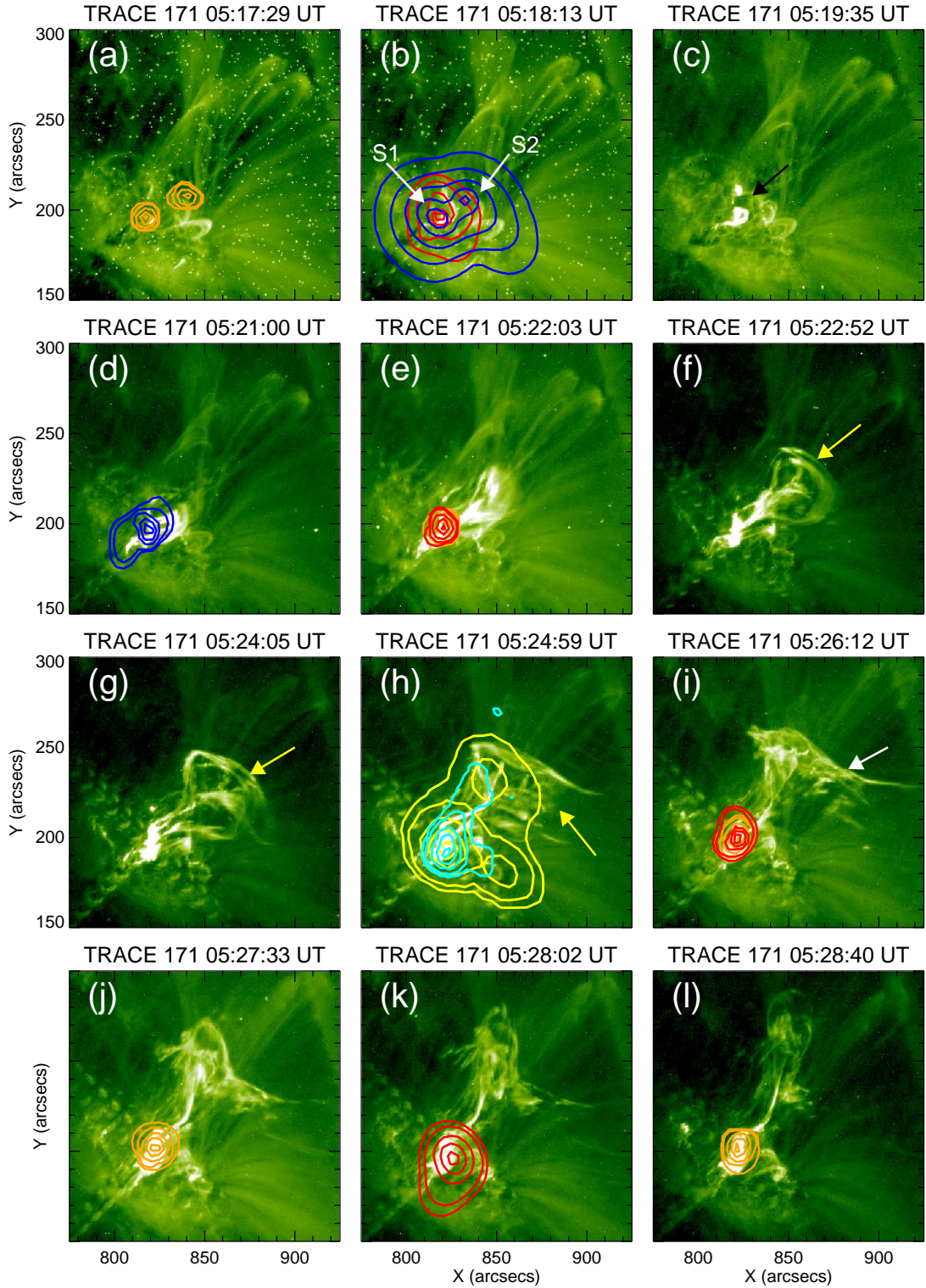}
\caption{Sequence of TRACE 171~\AA~images showing the activation, expansion and eruption of the prominence and associated flaring activity. The X-ray contours in different energy bands (12--25 (orange), 25--40 (red), 40--100 keV (blue)) are over plotted on some of the co-temporal EUV images. The conjugate emission centroids of HXR source at the 40--100 keV are marked by S1 and S2 in panel (b). For a comparison of the spatial location of HXR and MW emissions, we plot NoRH microwave images at 17 GHz (yellow) and 34 GHz (sky) in panel (h).  The contour levels for RHESSI and NoRH images are set as 5, 10, 30, 60, and 90\% of the peak flux of the each images.}
\label{MainEvent}
\end{figure}

\begin{figure}
\epsscale{1.0}
\plotone{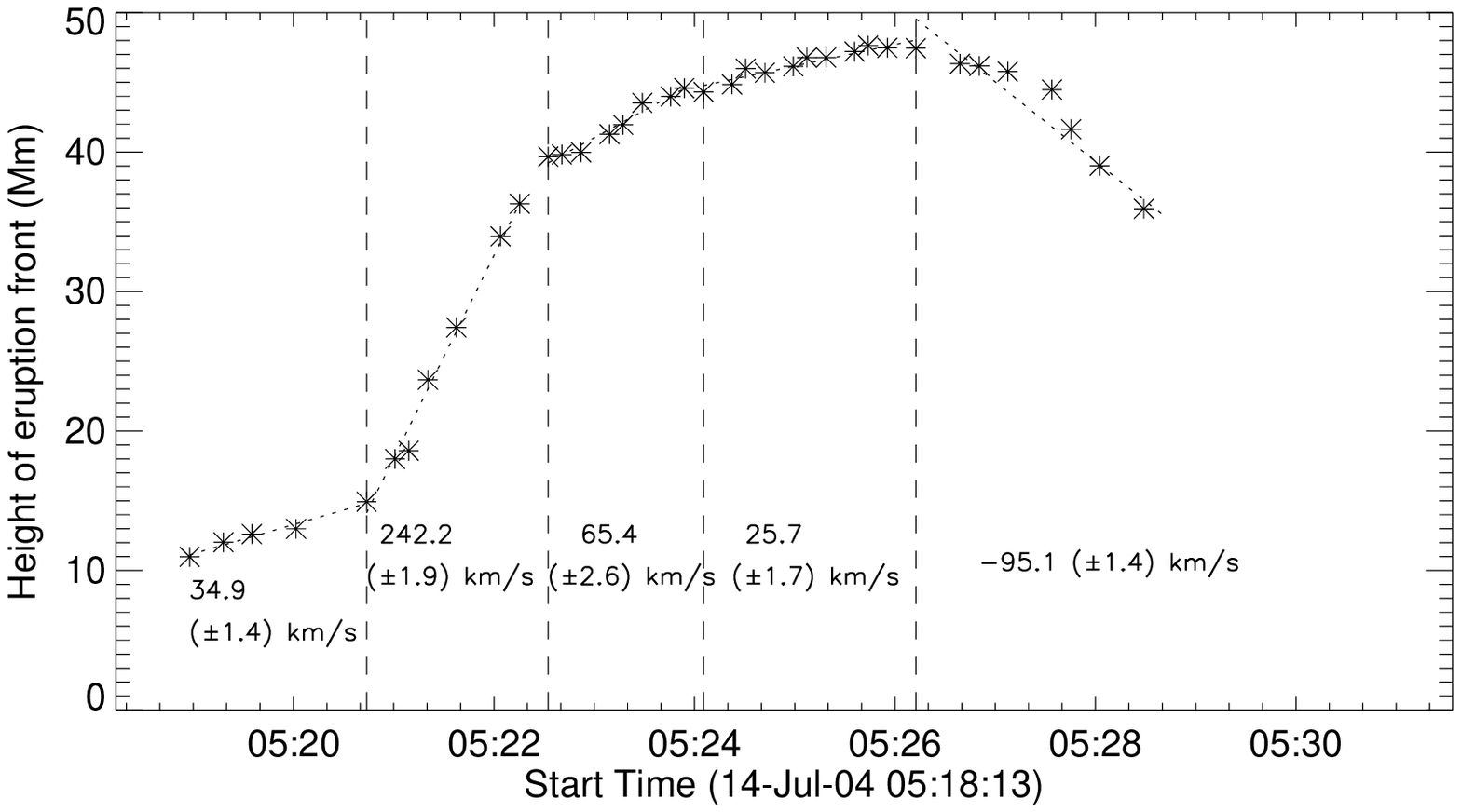}
\caption{Height-time plot of the eruption front during the main flare phase from 05:19 UT to 05:29 UT. The expansion of the eruption front shows different behaviour during different stages of eruption. In the early stage, the prominence started to rise with a constant velocity of $ \sim $35 km s$ ^{-1} $ till 05:20:40 UT. Thereafter, it suddenly accelerate to attain its maximum rising velocity of $ \sim $242 km s$ ^{-1} $ between 05:20:40 and 05:22:30 UT. After that the rising speed started slowed down from 242 to 65 and further, 65 to 26 km s$ ^{-1} $. Finally, the eruption front has completely halted at 05:26:10 UT after attaining a height of $ \sim $48 Mm above the solar surface and started downward motion with a speed of $ \sim $95 km s$^{-1}$.     }
\label{exp_ht}
\end{figure}

In Figure \ref{flux_profile}, we present the RHESSI and NoRH light curves along with the GOES time profiles for the M6.2 flare. We find that the HXR light curves $>$25 keV show a rapid increase in count rates from 05:17 UT onward indicating the onset of the impulsive phase of the flare. The non-thermal HXR emission continued till $\sim$5:25 UT which can be recognized in 25--40 keV light curve and further confirmed by HXR spectroscopy (see section~ \ref{sec:RHESSI_spec}). Therefore, we have considered the period from 05:17 to 05:25 UT as the impulsive phase of the flare which is indicated in Figure~\ref{flux_profile} by grey shaded area. The MW light curves clearly show that the impulsive phase is characterized by three distinct events of flux enhancement (marked as events `a', `b' and `c' in Figure~\ref{flux_profile}) with the first and third peaks at 05:18 UT and 05:21 UT being common in HXR profiles also. During MW peak `b' at 05:19:54 UT, the HXR profiles above 25 keV also exhibit flux enhancement but a distinct peak structure as in MW light curves can not be recognized. Further these MW peaks are much more distinguishable at 17 GHz than 34 GHz observations. After the peak `c', MW flux profiles (mainly 17 GHz) exhibit significant fluctuations during the impulsive phase.

\begin{figure}
\epsscale{0.9}
\plotone{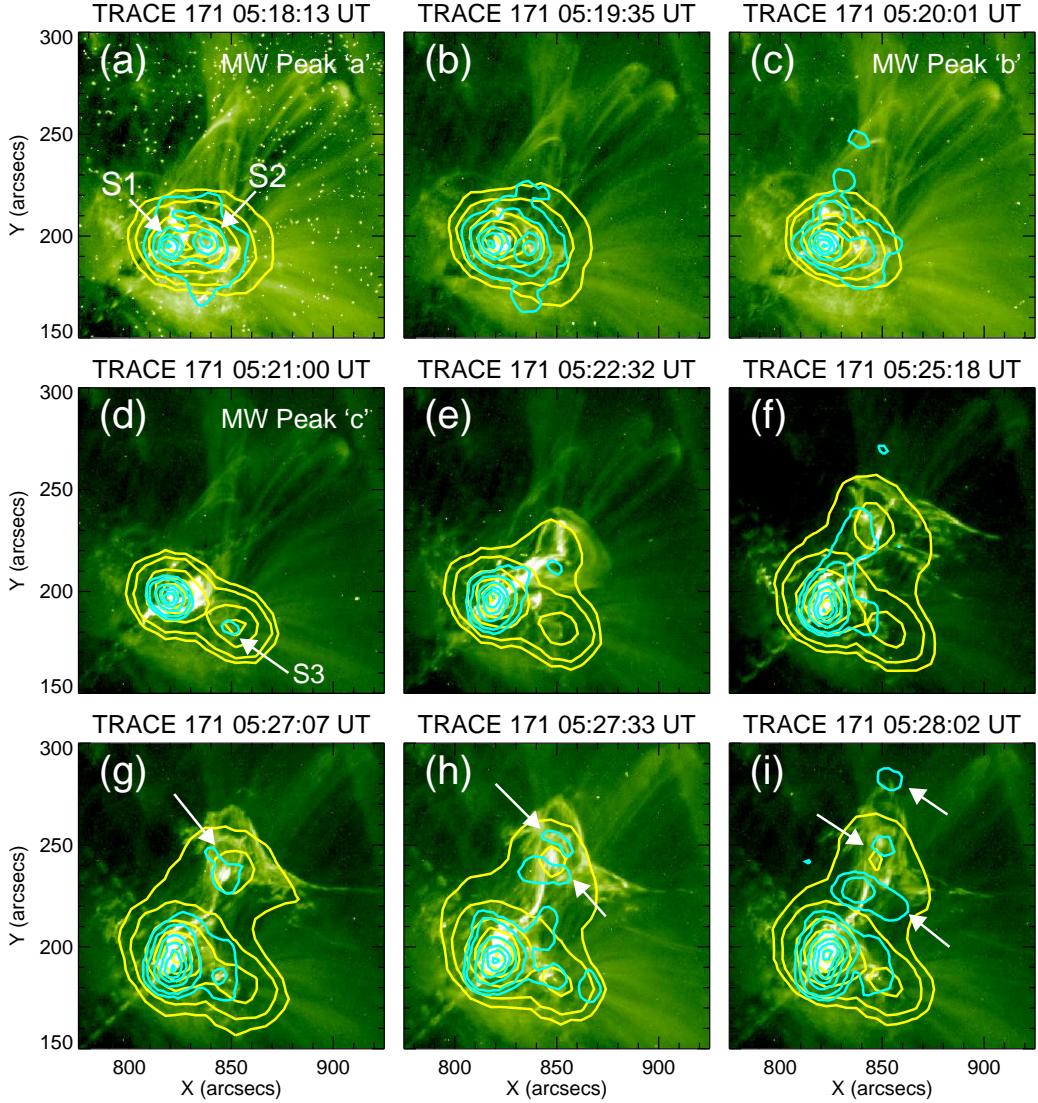}
\caption{Series of TRACE 171 \AA~images overlaid by co-temporal NoRH MW contours in 17 GHz (yellow) and 34 GHz (sky). The conjugate emission centroids of MW source at 34 GHz during early impulsive phase are marked by S1 and S2 in panel (a). We further indicate a relatively distant MW source by S3 in panel (d) which is more prominent at 17 GHz. The source S3 appeared during main impulsive peak (05:21 UT) and prevailed till the end of the flare ($ \sim $05:30 UT). The contour levels for MW images are set as 5, 10, 30, 55, 75, and 95\% of the peak flux of the each images. We note multiple coronal MW sources at both 17 GHz and 34 GHz and indicated them with arrows (white) in the panels (g), (h) and (i).}
\label{tr_norh}
\end{figure}

In Figure \ref{MainEvent}, we present combined imaging observations of the M6.2 flare in EUV, HXR, and MW to show the crucial phases of the flare evolution. Following the impulsive onset of HXR and MW emissions (event `a'), we observe the rise of the prominence at the source region and intense EUV brightening beneath it (Figures~\ref{MainEvent}(a)-(c)). In subsequent EUV images, we observe the evolution of filament in the corona with distinct appearance of its outermost structure (shown by arrow in Figures~\ref{MainEvent}(f), (g), and (i)) which we term as the `eruption front'. The eruption front likely represents a part of the magnetic structure associated with the erupting filament. In Figure \ref{exp_ht}, we provide height-time plot of the eruption front. The eruption front underwent fastest expansion between 05:21 and 05:23 UT with the speed of $\sim$242~km~s$^{-1}$. The eruption speed rapidly decreases after 05:23 UT. We also note that there is a further decrease in the speed of the eruption front as it moves to higher altitudes; speed of the eruption front was $ \sim $65 km s$ ^{-1} $ between $\sim$05:23--05:24 UT which decreased to $ \sim $26 km s$^{-1} $ between $\sim$05:24--05:26 UT. The eruption front was intact and symmetrical till $\sim$05:24 UT while attaining a height of 4.8$ \times $10$ ^{4} $ km (from the source region) at $\sim$05:25 UT. After 05:25 UT, the eruption front got disrupted (Figure \ref{MainEvent}(h)) while its major portion (right segment; marked by arrow in Figure \ref{MainEvent}(i)) descended which we could track/identify till $\sim $05:29 UT. In this interval, the downward speed of eruption front was estimated as $\sim$95~km~s$ ^{-1}$. The remaining portion of the eruption front narrowed down while moving further upward and eventually disrupted. In Figure~\ref{tr_norh}, we present a series of MW sources at 17 and 34 GHz over co-temporal EUV 171~\AA~images to study the spatial evolution of MW emission in relation to the phases of prominence activity.

LASCO on board  SOHO did not detect any CME associated with this flare\footnote{http:$//cdaw.gsfc.nasa.gov/CME\_list/UNIVERSAL/2004\_07/univ2004\_07.html$}. The absence of CME suggests that all the erupted material either fell down on the source region or arrested within the overlying coronal magnetic loops. 

\begin{figure}
\epsscale{1.0}
\plotone{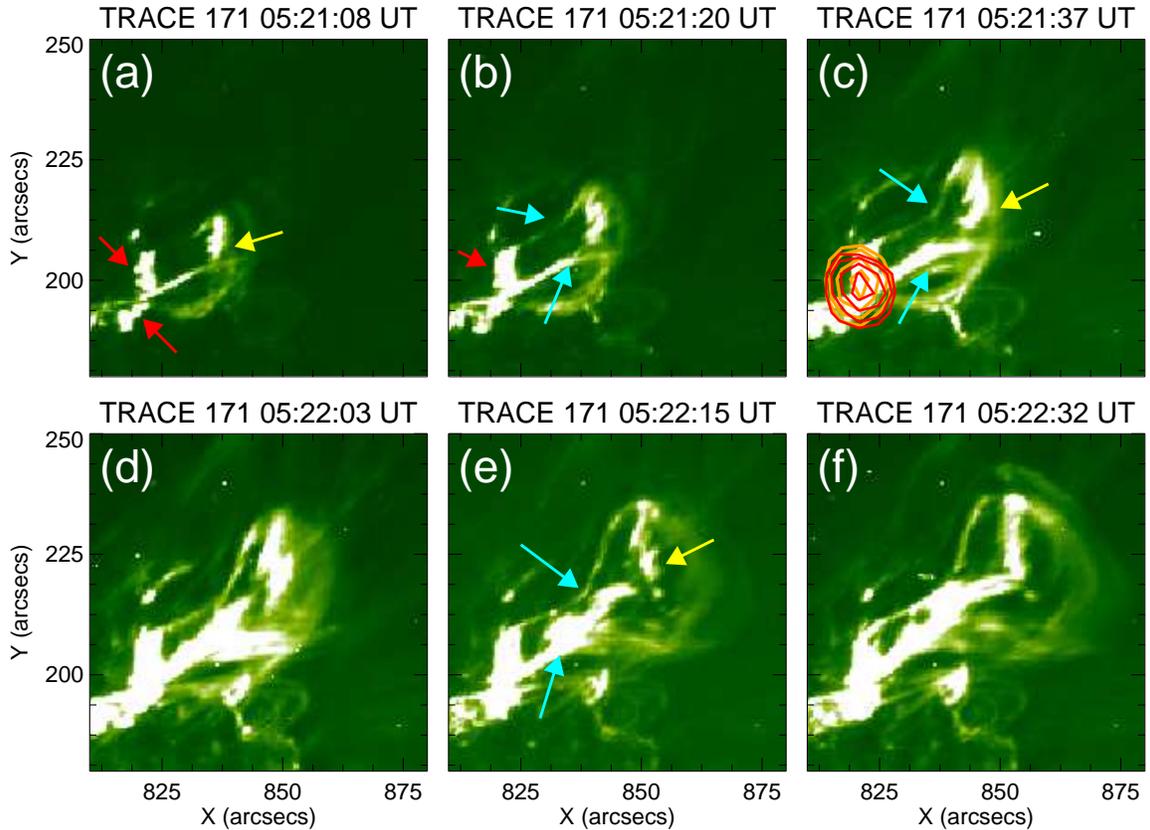}
\caption{Sequence of TRACE EUV images from 05:21--05:23 UT showing the fast expanding phase of the eruption. In this interval, we find a fast rising system of bright, twisted loops at the western side (yellow arrows) together with bright spiky structures at the eastern part (red arrows) that likely represent helically twisted loops associated with the erupting prominence. The X-ray contours in 12--25~keV (orange) and 25-40~keV (red) are shown in panel (c). The bending of magnetic field lines toward bright central region due to stretching by rising prominence-flux rope is indicated by arrows (sky) in panels (b), (c) and (e). We note that during this period the erupting system expanded with the maximum speed of $ \sim $242 km s$ ^{-1} $ (see Figure \ref{exp_ht}).}
\label{kink}
\end{figure}

EUV images from $\sim$05:21--05:23 UT showed a very important phase of eruption which requires more attention. We present enlarged view of EUV images in Figure \ref{kink} to emphasize some interesting morphological structures during the early evolution of the erupting system. We observe a system of twisted loops at the western side (leading side of eruption) that undergoes fast expansion (marked by yellow arrows). Further some bright spiky structures can be identified at the eastern side (marked by red arrows) which conceivably could be the projection of the helically twisted loops. To understand associated high energy emission, we present co-temporal HXR images at 12--25 (orange) and 25--40~keV (red) in Figure~\ref{kink}(c).

\subsubsection{RHESSI spectroscopy}          
\label{sec:RHESSI_spec}
                  
We have studied the evolution of RHESSI X-ray spectra during the flare over consecutive 20 s intervals from 05:17 UT to 05:25 UT. Further, we have analyzed spectra for a few selected intervals during the pre-flare peaks. For this analysis, we first generated a RHESSI spectrogram with an energy binning of 1/3 keV from 6 to 15 keV, 1 keV from 15 to 100 keV, and 5 keV from 100 to 200 keV energies. We only used the front segments of the detectors, and excluded detectors 2 and 7 (which have lower energy resolution and high threshold energies, respectively). The spectra were deconvolved with the full detector response matrix (i.e., off-diagonal elements were included; \citealt{Smith2002}). The time intervals for spectral analysis are carefully chosen to avoid the bad intervals due to changes in the attenuator states. In our case, there were two bad intervals during 05:18:04--05:18:08 (change in attenuator state from A$ _{0} $ to A$ _{1} $, i.e., thin shutter comes in) and 05:21:24--05:21:28 (change in attenuator state from A$ _{1} $ to A$ _{3} $, i.e., thin+thick (both) shutters are in).   

\begin{figure}
\epsscale{1.0}
\plotone{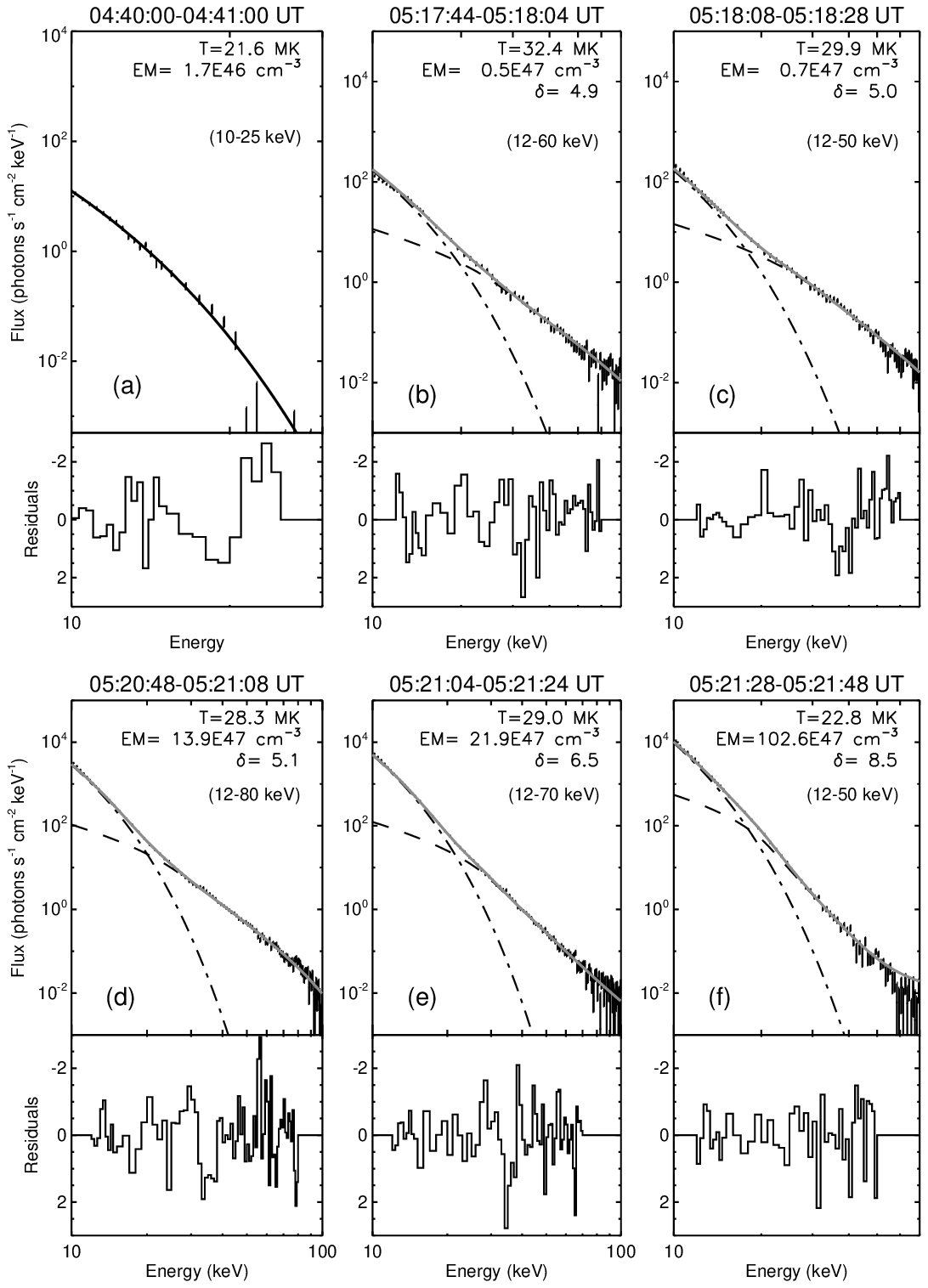}
\caption{RHESSI X-ray spectra along with their residuals derived during various time intervals before and during the M6.2 flare event on 2004 July 14. Panel (a) presents the spectrum which is fitted with an isothermal model during pre-flare peak I (defined in Figure \ref{goes_profile}). Panels (b)-(f) display a set of spectra that are fitted with a combination of isothermal model (dashed-dotted line) and thick-target bremsstrahlung model (dashed line). The grey (solid) line indicates the sum of the two components. The energy ranges chosen for fitting are shown on each panels.}
\label{hsi_spectra}
\end{figure}

In Figure \ref{hsi_spectra}, we illustrate spatially integrated, background subtracted RHESSI spectra along with their respective residuals for six selected intervals. Spectral fits were obtained using a forward-fitting method implemented in the OSPEX code. The HXR emission during pre-flare peaks was observed only up to 20 keV and can be well fitted with an isothermal model (see Figure \ref{hsi_spectra}(a)). During impulsive phase, we note HXR emission up to 100 keV. For this phase, spectral fittings were obtained using the bremsstrahlung and line spectrum of an isothermal plasma and a non-thermal thick-target bremsstrahlung model (i.e., combination of Vth and thick2 models; see Figures \ref{hsi_spectra}(b)-(f)). Also, we have chosen different time intervals for fitting during pre-flare peaks (1 minute) and impulsive phase (20 s) for better counts statistics. In Figure \ref{plasma_param}, we plot various spectroscopic parameters derived from spectral fittings: temperature ($T$) and emission measure ($EM$) from thermal fits; low-energy cutoff ($E _{LC} $) and electron spectral index ($ \delta $) from non-thermal thick-target fits. RHESSI observations reveal a very high plasma temperature ($T\sim $32 MK) at the time of first HXR peak at 05:18 UT (see Figure \ref{plasma_param}(a)) while $EM$ remains at lower values, which suggest intense heating within localized regions. Later on, both, the $T$ and $EM$ exhibit smooth variations (see Figures \ref{plasma_param}(a) and (b)).        

From the thermal parameters, we have further estimated density ($ n $) and pressure ($P$) of the flaring region (see Figure \ref{plasma_param}(c)). The density of the thermal plasma is inferred by the formula, $n$=$ \sqrt{EM/(f\cdot V)} $, where $EM$ is the emission measure, $V$ is the source volume, and $f$ is the volume filling factor. The volume has been estimated by the relation, $ V $=$ A^{3/2}$, where A is the area of X-ray sources at 50\% contour level of the corresponding 6--15 keV RHESSI images reconstructed with the CLEAN algorithm. The filling factor ($ f $), ratio of the volume filled by hot X-ray emitting plasma to the total source volume, is often assumed to be unity \citep[see, e.g.,][]{Sui2005}. The pressure of the plasma in thermal emission has been estimated by the formula,  $P$=$2n$\textit{k$ _{B} $}$T$, where \textit{k$ _{B} $} is Boltzmann's constant. It is noticeable that the density and pressure  show a slow yet steady rise until the second HXR peak at $ \sim $05:21 UT, but exhibit rather oscillatory behaviour in the later stages (see Figure \ref{plasma_param}(c)). 

The evolution of low-energy cutoff (E$_{LC} $) is shown in Figure \ref{plasma_param}(d). 
The E$_{LC} $ values derived are the upper estimates due to the dominance of the hot thermal contribution, and thus yield lower limits to the number of non-thermal electrons and their energies derived \citep{Holman2003}. The temporal evolution of electron spectral index ($ \delta $) clearly indicates spectral hardening at the two HXR peaks (at 05:18 and 05:21 UT) with electron spectral index ($ \delta $ $ \sim $5; see Figure \ref{plasma_param}(e) and spectra in Figures \ref{hsi_spectra}(b) and \ref{hsi_spectra}(d)) which suggest stronger acceleration of high energy particles at these epochs. 
\label{spectroscopy}

\subsubsection{Flare Energetics}

In the preceding subsection (\ref{spectroscopy}), we have estimated various spectroscopic parameters ($ EM $, $ T$, $ \delta $, $ F_{e}$, and $ E_{LC} $) which are used to evaluate thermal and non-thermal energies associated with the flare.
 
Using instantaneous values for $ T $ and $ EM $, the total thermal energy content can be estimated by following expressions
\begin{eqnarray}
E _{th} =3\textit{k$ _{B} $}TnV= 3\textit{k$ _{B} $}T\sqrt{EM\cdot f\cdot V}~[\text{erg}],
\end{eqnarray}

where \textit{k$ _{B} $, T, EM, V,} and \textit{f} are the Boltzmann constant, the plasma temperature (K), the emission measure (cm$ ^{-3} $), the source volume (cm$ ^{3} $) and volume filling factor, respectively. In accordance with previous RHESSI studies \citep[e.g.,][]{Holman2003, Veronig2005, Warmuth2013}, we assume unity for the filling factor ($f$). Recently \cite{Guo2012} determined from RHESSI imaging spectroscopy for a number of extended loop flares that the filling factor lies within the range 0.1 and 1 which they interpret as somewhat less than, but consistent with unity. 

\begin{figure}
\epsscale{0.90}
\plotone{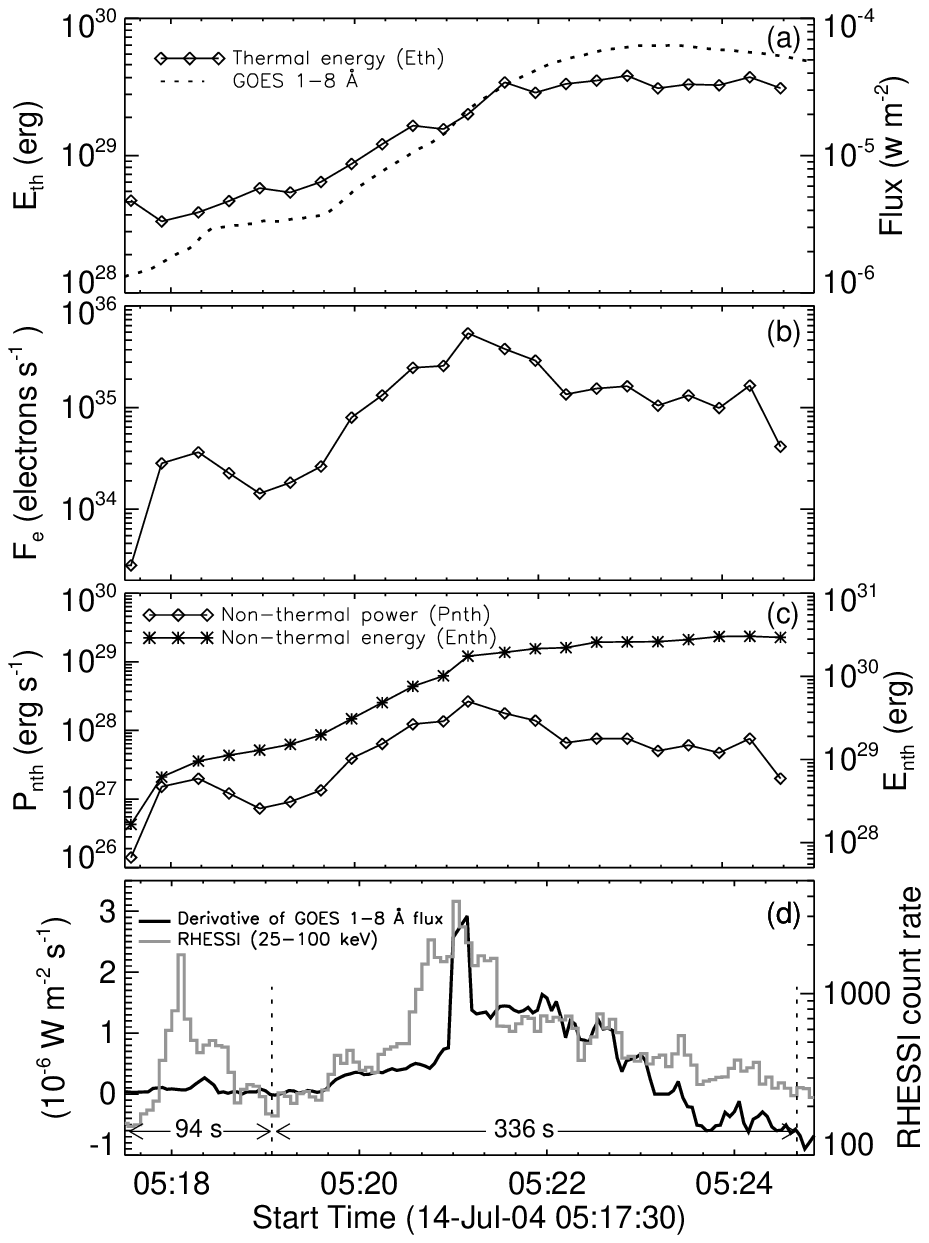}
\caption{Temporal evolution of (a) thermal energy (E$ _{th} $) along with GOES SXR flux at 1--8 \AA, (b) electron flux above 25 keV energy (F$ _{e} $), (c) non-thermal power (P$_{nth}$) and cumulative non-thermal energy (E$ _{nth} $) during the M6.2 flare. We have also shown the comparison of RHESSI HXR flux profile at 25--100 keV (grey) with derivative of GOES 1--8 \AA~flux (black) to check the Neupert effect in panel (d).}
\label{plasma_param2}
\end{figure}

The evolution of  $E_{th}$ is shown in Figure~\ref{plasma_param2}(a). We find that $E_{th}$ gradually builds up till the second HXR peak (05:21 UT) and remain roughly constant till the end of impulsive phase. $E _{th}$ maximizes $\sim$2 minutes after peak HXR flux and resembles the GOES SXR profiles (see Figure \ref{plasma_param2}(a)).      

\begin{figure}
\epsscale{0.75}
\plotone{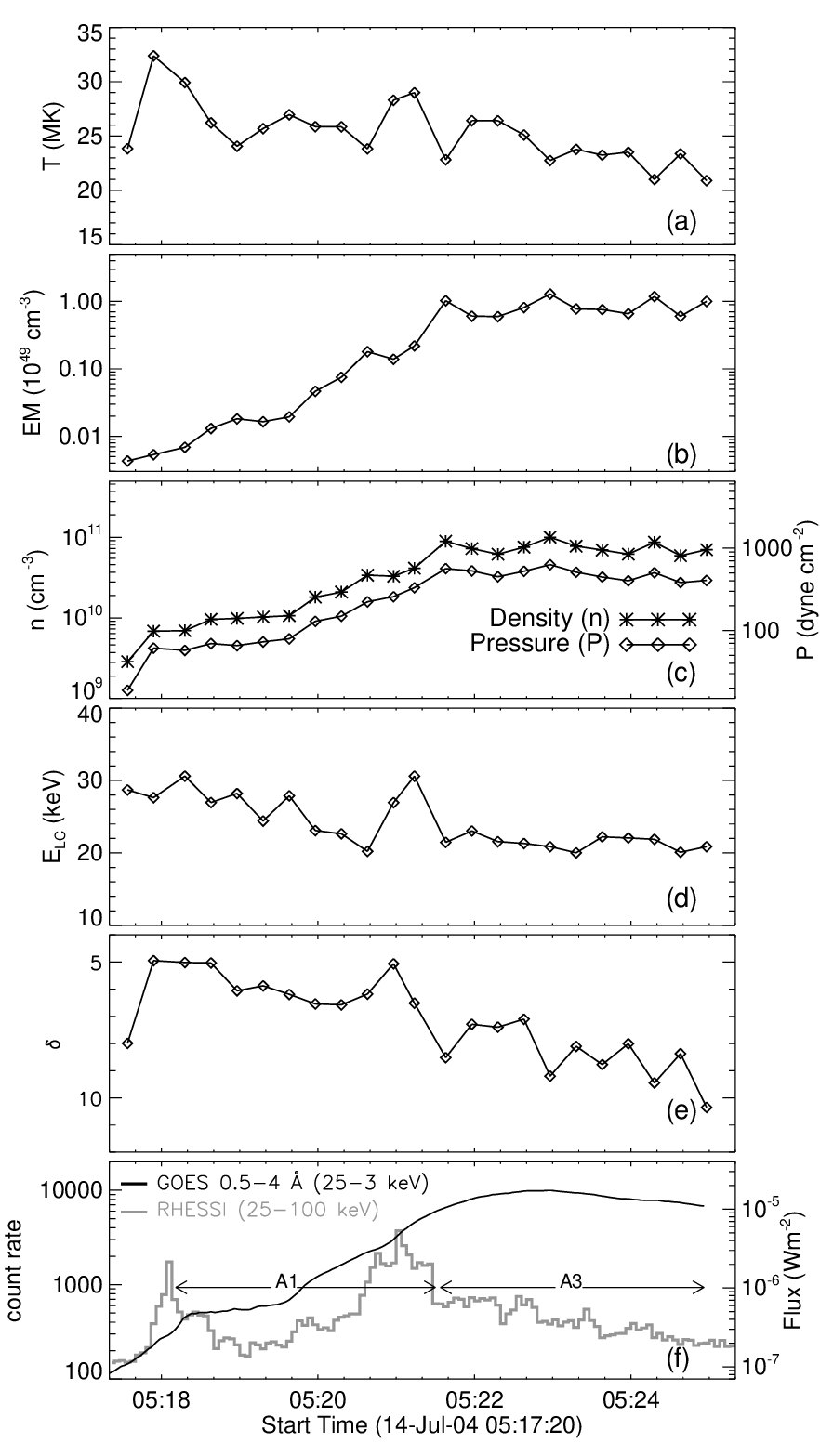}
\caption{Temporal evolution of various spectroscopic quantities derived from X-ray spectral analysis during the impulsive phase of the M6.2 flare. Panels (a)-(e): Temperature ($T$), emission measure ($EM$), density ($n$) along with pressure ($P$), low-cutoff energy (E$ _{LC} $), and electron spectral index ($ \delta $). To compare these parameters with respect to the flare evolution, we have plotted GOES high energy flux (0.4--5 \AA) and RHESSI 25--100 keV count rates in panel (f). We also indicated the time-duration in which the attenuators A1 and A3 were present in front of the RHESSI detectors.}
\label{plasma_param}
\end{figure}

In collisional thick-target model, the HXRs are produced by collisional bremsstrahlung during the passage of non-thermal electrons through denser plasma regions, in which the electrons are stopped completely by Coulomb collisions \citep{Brown1971,Brown2009,Kontar2011}. Using the electrons distribution parameters derived from RHESSI spectroscopy, the power delivered by non-thermal electrons above low-energy cutoff ($ E_{LC} $) can be calculated by the following expression 

\begin{eqnarray}
P_{nth}(E > E_{LC})=\frac{\delta-1}{\delta-2}F_{e}E_{LC}10^{35}~[\text{erg~s}^{-1}] ,
\label{eq2}
\end{eqnarray}
where $E _{LC} $ is the low-energy cutoff, $F _{e}$ is the total number of electrons per second above $E _{LC} $ in units of 10$ ^{35} $ electrons s$ ^{-1} $, and $\delta $ is the electron spectral index \citep{Fletcher2013}. An accurate determination of the low-energy cutoff to non-thermal electron distributions is crucial for the calculation of power and consequently non-thermal energy in solar flares \citep{Sui2005,Veronig2005}. In general flares are thought to have low-energy cutoffs close to or in the region where the emission is dominated by thermal bremsstrahlung \citep{Ireland2013}. We further emphasize that in flares with multiple HXR sub-peaks during the impulsive phase (like the present one), the determination of E$ _{LC} $ is rather illusive during the peak emission as it is difficult to distinguish the signals of flare accelerated electrons against dominant thermal bremsstrahlung. This often results in the overestimation of E$ _{LC} $ during HXR peak phases. 
For this event, we find that E$_{LC}$ varies in the range of 20-32~keV with relatively higher values ($\sim$32 keV) during the second peak (see Figure \ref{plasma_param}(d)). Therefore, in order to get an estimation of power (and consequently energy) of the non-thermal electrons, we have taken E$_{LC}$ as 25 keV which is the average of E$_{LC}$ over the flare time interval. The estimated non-thermal electron flux ($ F_{e} $) above 25 keV  is plotted in Figure \ref{plasma_param2}(b). In Figure \ref{plasma_param2}(c), we have plotted the power ($P _{nth} $) and energy (E$ _{nth}$) contained in non-thermal electron beams. The plots of $F _{e} $ and $P _{nth} $ clearly indicates noticeable enhancement in the particle rate and consequently power of flare-accelerated electrons during the two HXR peaks.          
       
\section{Discussion}
\label{discuss}

In this paper, we study coronal events that occurred in active region NOAA 10646 on 2004 July 14. We observed large-scale contraction of higher active region loops  for a time span of $ \sim $30 minutes which was followed by an M6.2 flare and associated failed eruption of a flux rope. Table~\ref{tab1} presents a summary of the different phases of flare evolution and associated phenomena.

\begin{table}
\begin{center}
\caption{Summary of various stages of flare evolution from pre-flare to decay phase.}
\begin{tabular}{p{1.6in}p{1.4in}p{3in}}
\tableline\\
Phases  & Time & Observations\\
\tableline\\
 
Pre-flare activity & 4:30 -- 05:17~UT  & Three small GOES flare events, EUV loop contraction for $ \sim $30 min, drastic change in speed of contraction during third peak of pre-flare phase, pre-flare events showing localized brightenings in EUV 171~\AA, MW and X-ray sources at the core region underneath the contracting coronal loops.  \\
Contraction phase & 04:47 -- 05:17 UT & Large overlying coronal loops observed at EUV 171~\AA~undergo contraction by $\sim$20 Mm (40\% of original height), loop contraction continues through all three episodes of pre-flare activity.\\
 
Flare impulsive phase and rapid activation of flux rope  & 05:17 -- 05:25 UT  & Peak HXR emission up to 100 keV with electron spectral index $\delta$ $\sim$5, multiple peaks in HXR and MW (17 and 34 GHz) emission, rapid rise of flux rope by $\sim$40 Mm. \\
Decay phase and confinement of eruption & 05:25 -- 05:35 UT & Braking and successive disruption of eruption front, multiple MW sources from footpoint and coronal regions, compact HXR footpoint source.\\
\tableline
\label{tab1}
\end{tabular}
\end{center}
\end{table}

\subsection{Contraction of coronal loops and pre-flare emission}
 
Although the active region NOAA 10646 had a relatively simple bipolar magnetic structure in the photosphere, it exhibited very complex structures of overlying loops in the corona (see Figure \ref{loops}). The EUV images at 171~\AA~clearly show large-scale contraction of a system of higher coronal loops that started $ \sim $30 minute prior to the M6.2 flare. It is important to note that the loop contraction was observed throughout the pre-flare phase and ceased with the onset of the flare impulsive phase. We observed three small events during the pre-flare phase (events I, II, and III; Figures \ref{goes_profile} and \ref{pre-2}) which are characterized by localized brightenings in the core region of the overlying loops. The speed of loop contraction increased drastically (from $ \sim $5 km s$ ^{-1} $ to $ \sim $25 km s$ ^{-1} $; Figure \ref{imp_ht}) just $ \sim $5 minute prior to the impulsive phase of the M-class flare, i.e., during the pre-flare event III. RHESSI measurements during the pre-flare events reveal HXR emission up to 20 keV. From the X-ray spectra, we find that the X-ray emission was predominantly thermal during this phase with plasma temperatures $ > $20 MK during pre-flare peaks (see e.g., Figure \ref{hsi_spectra}(a)). These observations indicate that the flaring site was already enveloped by hot plasma before the onset of filament activation and associated M6.2 flare. 

 It is worth to emphasize that the M6.2 flare occurred at the location of pre-flare brightenings which was enveloped by the large contracting coronal loop system. Although the phenomenon of loop contraction has been observed in several recent studies, the contraction reported here is remarkable in several aspects. First, the contraction was observed in overlying coronal loops at large-scales both at spatial and temporal domains during which the loop height decreased by $ \sim $20 Mm ($ \sim $40\% of original height). The total duration of loop contraction is $ \sim $30 minutes which is the longest period of loop contraction reported so far. Further, the contraction phase ends with the onset of the impulsive phase of the M6.2 flare, i.e., the onset of the impulsive phase can be treated as the transition from inward to outward motion of coronal loops. 
 
The investigations of loop contraction have emerged as a very important aspect of solar eruptions in recent times. These studies are essentially inspired by the RHESSI discovery of altitude decrease in the HXR loop top (LT) source during the early impulsive phase in SOL2002-04-15 of class M1.2 \citep{Sui2003}. This phenomenon of downward motion of HXR LT source was established by many subsequent RHESSI observations in flares of different intensity classes (i.e., from class C to X) during their earliest stages to the impulsive phase \citep{Sui2004, Liu2004, Veronig2006, Joshi2007, Liu2009, Joshi2009}. 
 
Motivated by the RHESSI observations, the dynamics of complex coronal loop system over the flaring core was extensively investigated in EUV images. These studies reveal a significant contraction of coronal loops that envelop the flaring region before and during the flare \citep[e.g.,][]{Li2006, Joshi2009, Gosain2012} which was attributed to `magnetic implosion' \citep{Hudson2000} by many authors \citep{Liu2009, Simoes2013}. \citet{Li2006} provided the first evidence for the contraction of EUV coronal loops that lasted for $ \sim $5 minutes from TRACE 195 \AA~images which was temporarily and spatially correlated with the shrinkage in RHESSI 12--25 keV LT source during the early impulsive phase of SOL2002-04-16 flare of class M2.5. The study of a long duration M7.6 event (SOL2003-10-24), characterized by a prolonged rise phase of $ \sim $20 minutes, provided one of the best examples for the contraction of coronal loops \citep{Joshi2009}. In this event, TRACE 195 \AA~loops and HXR LT sources underwent a shrinkage of $ \sim $35\% of initial height during $ \sim $11 minutes. More importantly, the downward motion was observed simultaneously with speed of $ \sim $15 km s$ ^{-1} $ in EUV loops as well as HXR LT sources in different energy bands, viz., 6--12, 12--25, and 25--50 keV.                

Due to observational limitations, it is usually not possible to probe the motions of flare associated LT and footpoint (FP) sources simultaneously. Nevertheless some uniquely observed events have given us the opportunity to examine the relationship between the dynamics of LT and FP sources in the early flare phases \citep{Ji2007,Liu2009,Liu2009a}. These studies indicate that shrinkage in LT sources are associated with converging FP motions, making us to predict that simultaneity of the two phenomena could be part of the coronal implosion. \citet{Liu2009} presented observations of contraction of large coronal loops during the early impulsive phase of SOL2005-07-30 of class C8.9. This contraction was found in three clusters of EUV loops observed at 171 \AA~that sustained for a longer interval with relatively slower speed ($ \sim $10 minutes with an average speed of $ \sim $5 km s$ ^{-1} $). Recently, \citet{Simoes2013} reported loop contraction in EUV 171 \AA~images during the impulsive phase of M6.4 flare SOL2012-03-09 that was associated with loop oscillations.

It is widely believed that coronal transients derive their energy from the energy stored locally in the coronal magnetic fields. \citet{Hudson2000} conjectured that the energy conversion process during transient events would involve a magnetic implosion when the following assumptions hold: (1) the energy required for the event must come from the corona directly; (2) gravitational potential energy plays no significant role; (3) low plasma $ \beta $ in the corona. With these assumptions, the conservation of energy implies that magnetic energy decreases between the static states before and after the energy release. The reduction of magnetic energy, $\int B^{2}/8\pi$ dV, and consequently the reduction of the magnetic pressure, B$ ^{2}/8\pi$, would inevitably result in the contraction of overlying field lines so as to achieve a new force balance. The observations of the contraction of overlying coronal loops during and before the impulsive phase of flares by high temporal cadence images from TRACE and SDO support the predictions of the conjecture by \citet{Hudson2000}. However, the observational evidences for the contraction in coronal loops are uncommon for the majority of flares that exhibit explosive rather than implosive behaviour.

We note that in our observations prolonged coronal implosion displays three distinct stages with varying speed of loop contraction (Figure \ref{imp_ht}(a)). 
Here it is important to focus on the brief and localized X-ray and MW emissions that were observed from the core region of the overlying loop system a few minute before the onset of contraction (Figures \ref{pre-2}(b)--(d)). Later on this region brightened up in EUV images also with the localized brightening of multiple low-lying loops. More importantly, the contraction speed rapidly increased $ \sim $5 minute before the flare onset and this phase is co-temporal with the sequential brightenings of low-lying coronal loops at the core region during which X-ray and MW emissions intensified (Figure \ref{pre-2}). The observations of sequential and localized brightenings during pre-flare phase imply episodic release of small amount of energy which can be considered as the signature of magnetic reconnection at relatively small-scales (in contrast to large-scale reconnection during the flare impulsive phase) \citep{Chifor2007, Joshi2011, Joshi2013, Awasthi2014}. RHESSI observations of this phase further indicate that this interval is mostly dominated by the thermal emission. The slow and gradual energy release over the prolonged pre-flare phase will cause the magnetic pressure of the coronal loop system to decrease. It is likely that during this slow preheating phase, the increase of the thermal pressure is mainly due to localized heating and gentle evaporation which is not enough to compensate for the decrease of the magnetic pressure \citep[see][]{Liu2009}. This will result in the implosion of the surrounding flaring region consisting of EUV coronal loops. We therefore conjecture that multiple small-scale, localized events of energy release at the core region over the prolonged preheating phase favourably contributed to sustain a large-scale contraction of coronal loops. With the onset of the `standard flare reconnection' during the impulsive energy release, the implosion becomes insignificant as the large-scale magnetic reconnection proceeds successively in higher coronal loops stretched by the eruption of the prominence or magnetic flux rope.

\subsection{Multi-wavelength flare emissions and confinement of the erupting prominence}

We examine the temporal and morphological evolution of MW and HXR emission during the impulsive phase (i.e., shaded area in Figure~\ref{flux_profile}) and their associations with the prominence activation. The comparisons of spatial and temporal structures of HXR and MW emission offer us crucial information on different aspects of conditions in the solar atmosphere where the flare occurs as both types of emissions are produced by highly energetic electrons but with different emission mechanisms. The HXR emission most likely arises from bremsstrahlung produced when energetic electrons are decelerated by Coulomb forces in collisions with the ambient ions, either in the chromosphere or in the corona. Bremsstrahlung HXR emission is proportional to the product of the non-thermal electron density and the ambient ion density. On the other hand, non-thermal MW emission is produced by the gyrosynchrotron mechanism which depends on the magnetic field intensity and its direction. The gyrosynchrotron production mechanism is very efficient and allows us to detect electrons at energies of hundreds of keV, even when their numbers are very low. Microwave emission can also be produced by thermal electrons through bremsstrahlung in sufficiently dense thermal plasma \citep[for a review see][]{White2011}.

The impulsive phase is characterized by three events of flux enhancements in MW emission denoted by `a', `b', and `c'. It is worth to note that the events `a' and `c' occurred simultaneously in MW and HXR profiles while `b' is largely missing in HXR measurements. The spatial evolution of HXR emission with respect to EUV images (see Figure~\ref{MainEvent}) provides some important insights to understand the relationship between flare emission and phases of prominence activity. We find that at the very beginning of the impulsive phase ($\sim$5:17 UT; Figure~\ref{MainEvent}(a)), two distinct 12--25 keV HXR sources are observed and at this time the flare related EUV brightening at HXR source locations is insignificant.  Soon afterwards, non-thermal high energy HXR emission (at 40$-$100 keV energy band) was observed from the same location with two distinct, well separated emission centroids (see Figure~\ref{MainEvent}(b)). This phase corresponds to the first peak of the impulsive phase (i.e., event `a' marked in Figure~\ref{flux_profile}). At this time, plasma temperature (estimated from RHESSI spectra; Figure~\ref{hsi_spectra}(b)) raised to $\sim$32~MK~which also corresponds to the maximum temperature during the flare (Figure~\ref{plasma_param}(a)). The high temperature and low emission measure (Figure~\ref{plasma_param}(a) and (b)) indicate impulsive heating of plasma within a small volume which is also consistent with corresponding EUV images that show localized brightening. We also note that this very epoch is associated with significant non-thermal characteristics with an electron spectral index $\delta$ $\sim$5 (Figures~\ref{hsi_spectra} and \ref{plasma_param}). It is further noteworthy that during event `a', high energy HXR source at 40$-$100~keV resembles the MW source at 34 GHz; both show emission from an extended region with the appearance of two distinct well separated emission centroids (see also Figures~\ref{MainEvent}(b) and \ref{tr_norh}(a)). In Figure \ref{mdi_hsi}, these conjugate emission centroids are indicated on combined WL/magnetogram image as S1 and S2. The associations of S1 and S2 with opposite polarity magnetic regions suggest that they probably represent emissions from the footpoints of a flaring loop. On the other hand the 17 GHz source resembles the HXR emitting structure at relatively lower energies ($<$40 keV). We also note that event `a' is followed by the rise and subsequent eruption of the prominence.

The peak of the impulsive phase (i.e., event `c') occurred simultaneously in HXR and MW emission. At this time, HXR sources at energies $>$25~keV~exhibit a single structure (Figure~\ref{MainEvent}(d)). The non-thermal HXR emission is generally believed to originate from footpoint locations of coronal loops by thick-target bremsstrahlung, thus observations of pairs of HXR sources are expected. The single HXR source implies that the two footpoints are very close indicating energy release in small low-lying loops \citep[e.g.,][]{Kushwaha2014}. Here it should be noted that the present event occurred close to the limb, so due to projection effects the footpoint sources may appear much closer than their actual separation. Although this HXR peak (event `c') is much broader and intense than the first peak (event `a'), the non-thermal electron spectral indices during the two peaks are comparable ($\sim$5; see Figures~\ref{hsi_spectra} and \ref{plasma_param}). 

\begin{table}
\begin{center}
\caption{Important characteristics of impulsive phase of the M6.2 flare 
derived from RHESSI spectroscopy (refer to Figure \ref{plasma_param2}).}
\begin{tabular}{p{3.2in}p{2.0in}}
\tableline\\

Flare characteristics  & Parameters\\
\tableline\\

Total duration of HXR peaks & 430 s \\
No. of HXR peaks  & 2\\
            & 94 s and 336 s \\
Total non-thermal energy ((E$_{nth}$)$ _{tot} $)   &     3.03  $\times $10$ ^{30} $erg\\
Thermal energy (E$_{th} $)  & \\
     ~~~~~~~~~~~~~~~ --(E$_{th}$)$ _{max} $               & 3.89  $\times $10$ ^{29} $ erg\\
     ~~~~~~~~~~~~~~~ --(E$_{th}$)$ _{min} $  &              0.33 $ \times $10$ ^{29} $ erg\\

$({E_{nth}})_{tot}/({E_{th} })_{max}$ &  $ \sim $7.5\\             

\tableline
\label{tab2}
\end{tabular}
\end{center}
\end{table}

In table \ref{tab2}, we summarize various aspects of energy release during the impulsive phase of the flare (see also Figure \ref{plasma_param2}). The total duration of HXR impulsive phase (as revealed by HXR flux profile $ > $25 keV) is 430 s which is composed of two peaks of 94 s and 336 s durations (see Figure \ref{plasma_param2}(d)). We find that the power delivered by the non-thermal electrons (P$_{nth} $) was maximum during the peak of HXR impulsive phase (event `c') while the thermal energy E$_{th} $ maximized $ \sim $2 min later which nearly coincides with the peak of thermal SXR emission (Figures \ref{plasma_param2}(a) and (c)). The total non-thermal energy ((E$_{nth} $)$ _{tot} $) emitted during this whole impulsive phase was $ \sim $3.0$ \times $10$ ^{30} $erg while the maximum thermal energy ((E$ _{th} $)$ _{max} $) was noted at the end of impulsive phase as $ \sim $3.9$ \times $10$ ^{29} $ergs, yielding a ratio of (E$ _{nth} $)$ _{tot} $/(E$ _{th} $)$ _{max} $ $ \sim $7.5. \citet{Saint-Hilaire2005} examined the ratio of non-thermal to thermal energies for a set of C- and M-class flares and found that this ratio varies in the range of $ \sim $1.5--6. It was also noted the flares with longer HXR peaks ($>$200 s) display larger ratios of non-thermal to thermal energy. Considering that the present event displays a longer HXR impulsive phase of 430 s with two sequential HXR peaks, the ratio of $ \sim $7.5 is in good agreement with the results of \citet{Saint-Hilaire2005}. 

In order to understand the relationship between the flare-accelerated electron beam and the flare-associated emission from thermal plasma, we compare the HXR profile (RHESSI 25--100 keV) with the derivative of SXR flux (GOES 1--8 \AA) in Figure \ref{plasma_param2}(d). We obtained a good temporal consistency between the two curves which suggests that this flare, associated with confined eruption, follows the Neupert effect \citep{Hudson1991,Dennis1993,Veronig2005}. The consistency with the Neupert effect is further complemented by the fact that the curve showing the evolution of cumulative non-thermal energy (Figure \ref{plasma_param2}(c)) exhibits a good temporal correspondence with the evolution of thermal energy (Figure \ref{plasma_param2}(a)) indicating that the non-thermal energy is eventually converted into the thermal energy \citep{Saint-Hilaire2005}. The validity of the Neupert effect implies that the energetic electrons responsible for the HXR emission by thick-target collisional bremsstrahlung (imaged as high energy HXR sources) are the main source of heating and mass supply of the SXR emitting hot coronal plasma \citep{Veronig2005}. We also note that with the onset of prominence activation and increase of flare emission after event `a', there is an enhancement in the density and pressure of the thermal plasma (Figure~\ref{plasma_param}(c)). 

With the onset of event `c', we observe a rapid increase in the speed of the erupting prominence (Figure~\ref{exp_ht}). It is important to note that HXR images at high energies ($>$25~keV) do not show significant changes in the location and structure of the HXR source, i.e., a single compact source is observed that remained at the same location. We note that the eruption proceeded symmetrically till $ \sim $05:24 UT with the distinct appearance of an intact eruption front (shown by arrow in Figure \ref{MainEvent}(f) and (g)). Thereafter, the eruption front evolved into an asymmetrical structure (first seen in EUV image at 05:24:30 UT) with the disruption of the south-west part of the eruption front (marked by arrow in Figure~\ref{MainEvent}(h)). In the successive images, we clearly observe the downfall of the erupted material which eventually leads to the complete failure of the filament eruption without any CME. \cite{Mrozek2011} and \cite{Netzel2012} have carried out detailed investigations to understand the reason for the confinement of the prominence for this event. They found strong evidence that the interaction between overlying coronal magnetic fields with the erupting prominence was capable to suppress the eruption completely. 

We have recognized some interesting morphological structures associated with the erupting prominence that overlap the flare's impulsive phase ($\sim$05:21-05:23~UT) and highlighted in EUV~171~\AA~images in Figure~\ref{kink}. Here we emphasize that this phase corresponds to the interval when the erupting prominence attained the maximum speed of $\sim$242~km~s$^{-1}$ (Figure~\ref{exp_ht}). These structures consists of rapidly expanding bright twisted loops at the front and spiky patterns at the following part which very likely represent the helically twisted loops associated with the erupting prominence. These structures can be attributed to the portions of the heated prominence within the flux rope. The HXR source at 25--40 keV along with corresponding spectrum (Figure~\ref{hsi_spectra}(f)) presents evidence for non-thermal emission from accelerated electrons in that region in addition to hot plasma. Further, from the EUV images, it is apparent that the magnetic field lines underwent curving toward the bright central region after being stretched by the erupting prominence-flux rope system (shown by arrows in Figures~\ref{kink}(b), (c), and (e)). 

Flux ropes are considered to be an important structural component in the models of solar eruptions. Many recent studies validate the existence of flux ropes in the lower corona in EUV observations, mostly in hot EUV channels \citep{LiJ2005, Cheng2011, Kumar2012, Patsourakos2013, JoshiNC2014, Vema2014}. Here we emphasize that evidence for a flux rope was found just after the peak `c' (the largest peak) of the flare impulsive phase (Figure~\ref{flux_profile}). It is probable that the flux rope was formed following this most violent episode of energy release. \cite{LiJ2005} presented a very clear example of formation and rise of an EUV flux rope structure during the impulsive phase of an X-class flare. These observations suggest that a flux rope has a multi-temperature structure that possesses several structural components, such as, sigmoid, hot plasma blob or plasmoid, leading edge, etc. In our observations, the flux rope structure was seen in EUV 171~\AA~ (which represents plasma at a temperature of $\sim$1 MK) presumably due to the heating of the prominence that lies within the flux rope. \cite{Gilbert2007} illustrated three kinds of magnetic topology that can lead to the diversity of eruptive phenomena: full, partial and nil \citep{Gilbert2001}. According to this study, magnetic reconnection can occur completely above the system of a prominence and its supporting flux rope or within it leading to nil or partial eruption of the filament. We believe that, the eruption of bright prominence material along with the helically twisted loops represents the case of magnetic reconnection occurring within the system of prominence and supporting magnetic flux rope. However, we believe that this is just a part of the flux rope which contain the prominence body. It is very likely that the remaining part of the flux rope, devoided of the prominence, might exist at a very high temperature which is not visible in 171~\AA~EUV images. 

MW sources at 17 and 34 GHz show an interesting evolution from lower as well as higher coronal regions (Figure \ref{tr_norh}). In general, we note significant differences in the morphology of MW and HXR sources although one of the MW sources (which appeared at the early impulsive phase; see Figure \ref{MainEvent}(h)) exhibits spatial correlation with the HXR source throughout the flare. During event `a' (i.e., between 05:18--05:19 UT), we observe a 34 GHz MW source with two distinct centroids having a separation of $ \sim $18$ \arcsec $ (see Figure \ref{tr_norh}(a) and (b)). On the other hand, 17 GHz images (note that 17 GHz images have lower spatial resolution than 34 GHz images) show a relatively extended single source.  As discussed earlier, the conjugate centroids (S1 and S2; Figures \ref{mdi_hsi} and \ref{tr_norh}(a)), observed at high energy HXR (40$-$100 keV) and 34 GHz MW emissions, exhibit structural similarities and co-spatiality during event `a'. At the second peak (event `b'; see Figure \ref{flux_profile}), images at both MW frequencies present similar source structures with a single centroid (see Figure \ref{tr_norh}(c)). During the third MW burst at 05:21 UT (event `c'; Figure \ref{flux_profile}), a new MW source originated at a distance of $\sim $33$ \arcsec$ toward the south-west of the pre-existing source (see Figure \ref{tr_norh}(d)-(i)). We mark the newly developed source as S3 (Figure \ref{tr_norh}(d)) and show its location on WL/magnetogram in Figure \ref{mdi_hsi}. This new source S3 prevailed till the end of the flare ($ \sim $05:30 UT) with more prominent appearance at 17 GHz over 34 GHz. We also note the source S3 never appeared in HXRs. The temporal and spatial associations of HXR and MW sources (also see Figure \ref{mdi_hsi}) indicate that the main energy release site lies in coronal loops close to the magnetic neutral line, probably formed by connecting footpoints S1 and S2. It is also likely that the distant MW source S3 resulted from the injection of flare accelerated electrons onto an overlying coronal loop (which probably connects S1 and S3 regions) as the rising flux rope interacted with overlying loop systems. We note that NoRH has much better dynamic range than RHESSI and therefore it is capable of detecting secondary sources \citep{White2011}. 

More importantly, we find multiple MW sources from higher coronal regions that intermittently appear at several locations after $\sim$05:22 UT (Figures \ref{tr_norh}(e)--(i)). It is also important to note that the whole flaring region (i.e., footpoint and coronal regions associated with the prominence eruption and its subsequent confinement) brightened up in 17 GHz MW emission following the impulsive phase of the flare. In particular, distinct MW sources appeared along the narrow, bright region seen in EUV images where bright blobs of plasma were observed following the confinement of the eruption (marked by arrows in Figures~\ref{tr_norh}(g)-(i)). These multiple coronal MW sources were observed after the beginning of disruption of the eruption front, i.e., when the ejected prominence and its supporting flux rope were subjected to confinement by the overlying field lines. In view of this, we conclude that the distinct MW sources in the corona presumably represent emission from hot plasma blobs formed within the collapsing magnetic flux rope.

\section{Summary and conclusions}
We have presented a comprehensive multi-wavelength study of an M6.2 flare which is associated with a confined eruption of a prominence-flux rope system. The availability of a wide range of multi-wavelength data (EUV, MW, X-ray) from the beginning of the pre-flare to the end of the impulsive phase of the event provided us with a unique opportunity to probe several flare associated phenomena in detail: large-scale loop contraction, small-scale loop brightenings, eruption of prominence-flux rope system and its subsequent confinement. In the following, we summarize the important results of this study:
\begin{enumerate}
\item
We have reported implosion of overlying coronal loops that continued over 30 minute during which overlying loops underwent contraction by 20 Mm ($ \sim $40 \% of their original height) during the pre-flare phase. Such a large-scale contraction has been reported for the first time. We observe episodic and localized events of energy release in low-lying loops at the core of the large overlying loops.  By synthesizing the multi-wavelength data, we propose that prolonged loop contraction is a manifestation of localized events of energy release that occurred intermittently during the pre-flare phase of the M6.2 flare.
\item
The pre-flare phase was dominated by thermal emission with temperatures beyond 20 MK. This indicates that the plasma was already substantially preheated at the flare core before the onset of impulsive phase. We believe that the strong preheating at the flare core will contribute favorably toward efficient particle acceleration during the subsequent impulsive phase of the event.  

\item
The impulsive phase of the flare is characterized by multiple non-thermal peaks. After the first impulsive peak, associated with strong HXR emission up to 40-100 keV, we observe the activation of the prominence and its supporting flux rope. Our observations imply that the onset of impulsive flare emission (which probably corresponds to large-scale magnetic reconnection) has triggered the eruption of the flux rope. 
\item 

RHESSI spectroscopy reveals high plasma temperatures ($ \sim $30 MK) and substantial non-thermal characteristics with electron spectral index ($\delta \sim$5) during the impulsive phase of the flare. During this phase, characterized by two sequential HXR peaks, the ratio of total non-thermal energy to maximum thermal energy was found to be $ \sim $7.6 which is consistent with earlier studies. More importantly, the time-evolution of thermal energy nicely correlates with the variations of the cumulative non-thermal energy throughout the impulsive phase of the flare. This can be interpreted in terms of conversion of the energy of accelerated particles to hot flare plasma and is well consistent with the Neupert effect.  

\item
The prominence along with its supporting flux rope could not have a successful escape through the overlying coronal loops and therefore leads to a confined eruption. The observations of the confinement process of the flux rope are remarkable; we detect multiple coronal MW sources along the trajectory of the eruption. The EUV images show hot plasma blobs on the location of these coronal sources. In our opinion, the coronal MW sources represent emission from hot plasma blobs which are formed within the collapsing magnetic flux rope as a result of its interaction with the overlying and surrounding magnetic field lines. 
\end{enumerate}

This paper highlights the importance of studying the pre-flare activity. This study also underlines that confined eruptions form a very interesting category of solar eruptive phenomena. Their investigations provide a unique opportunity to probe the interaction among different magnetic field systems in the corona besides exploring the triggering mechanisms and energy releases processes.

\acknowledgments
We acknowledge RHESSI, TRACE, NoRH, SOHO, and GOES for their open data policy. RHESSI and TRACE are NASA's Small Explorer Missions. SOHO is a joint project of international cooperation between the ESA and NASA. This work was supported by the BK21 plus program through the National Research Foundation (NRF) funded by the Ministry of Education of Korea. A.M.V. gratefully acknowledges the Austrian Science Fund (FWF): P27292-N20. We sincerely thank the anonymous referee for providing constructive comments and valuable suggestions that have enhanced the quality and presentation of this paper. 


\begin{thebibliography}{}
\expandafter\ifx\csname natexlab\endcsname\relax\def\natexlab#1{#1}\fi

\bibitem[{{Alexander} {et~al.}(2006){Alexander}, {Liu}, \&
  {Gilbert}}]{Alexander2006}
{Alexander}, D., {Liu}, R., \& {Gilbert}, H.~R. 2006, \apj, 653, 719

\bibitem[{{Alexander} \& {Metcalf}(1997)}]{Alexander1997}
{Alexander}, D., \& {Metcalf}, T.~R. 1997, \apj, 489, 442

\bibitem[{{Awasthi} {et~al.}(2014){Awasthi}, {Jain}, {Gadhiya}, {Aschwanden},
  {Uddin}, {Srivastava}, {Chandra}, {Gopalswamy}, {Nitta}, {Yashiro},
  {Manoharan}, {Choudhary}, {Joshi}, {Dwivedi}, \& {Mahalakshmi}}]{Awasthi2014}
{Awasthi}, A.~K., {Jain}, R., {Gadhiya}, P.~D., {et~al.} 2014, \mnras, 437,
  2249

\bibitem[{{Benz}(2008)}]{Benz2008}
{Benz}, A.~O. 2008, Living Reviews in Solar Physics, 5, 1

\bibitem[{{Brown}(1971)}]{Brown1971}
{Brown}, J.~C. 1971, \solphys, 18, 489

\bibitem[{{Brown} {et~al.}(2009){Brown}, {Turkmani}, {Kontar}, {MacKinnon}, \&
  {Vlahos}}]{Brown2009}
{Brown}, J.~C., {Turkmani}, R., {Kontar}, E.~P., {MacKinnon}, A.~L., \&
  {Vlahos}, L. 2009, \aap, 508, 993

\bibitem[{{Carmichael}(1964)}]{Charmichael1964}
{Carmichael}, H. 1964, NASA Special Publication, 50, 451

\bibitem[{{Cheng} {et~al.}(2011){Cheng}, {Zhang}, {Liu}, \& {Ding}}]{Cheng2011}
{Cheng}, X., {Zhang}, J., {Liu}, Y., \& {Ding}, M.~D. 2011, \apjl, 732, L25

\bibitem[{{Chifor} {et~al.}(2006){Chifor}, {Mason}, {Tripathi}, {Isobe}, \&
  {Asai}}]{Chifor2006}
{Chifor}, C., {Mason}, H.~E., {Tripathi}, D., {Isobe}, H., \& {Asai}, A. 2006,
  \aap, 458, 965

\bibitem[{{Chifor} {et~al.}(2007){Chifor}, {Tripathi}, {Mason}, \&
  {Dennis}}]{Chifor2007}
{Chifor}, C., {Tripathi}, D., {Mason}, H.~E., \& {Dennis}, B.~R. 2007, \aap,
  472, 967

\bibitem[{{Dennis} \& {Zarro}(1993)}]{Dennis1993}
{Dennis}, B.~R., \& {Zarro}, D.~M. 1993, \solphys, 146, 177

\bibitem[{{F{\'a}rn{\'{\i}}k} {et~al.}(2003){F{\'a}rn{\'{\i}}k}, {Hudson},
  {Karlick{\'y}}, \& {Kosugi}}]{Farnik2003}
{F{\'a}rn{\'{\i}}k}, F., {Hudson}, H.~S., {Karlick{\'y}}, M., \& {Kosugi}, T.
  2003, \aap, 399, 1159

\bibitem[{{Fletcher} {et~al.}(2013){Fletcher}, {Hannah}, {Hudson}, \&
  {Innes}}]{Fletcher2013}
{Fletcher}, L., {Hannah}, I.~G., {Hudson}, H.~S., \& {Innes}, D.~E. 2013, \apj,
  771, 104

\bibitem[{{Fletcher} {et~al.}(2001){Fletcher}, {Metcalf}, {Alexander}, {Brown},
  \& {Ryder}}]{Fletcher2001}
{Fletcher}, L., {Metcalf}, T.~R., {Alexander}, D., {Brown}, D.~S., \& {Ryder},
  L.~A. 2001, \apj, 554, 451

\bibitem[{{Fletcher} {et~al.}(2011){Fletcher}, {Dennis}, {Hudson}, {Krucker},
  {Phillips}, {Veronig}, {Battaglia}, {Bone}, {Caspi}, {Chen}, {Gallagher},
  {Grigis}, {Ji}, {Liu}, {Milligan}, \& {Temmer}}]{Fletcher2011}
{Fletcher}, L., {Dennis}, B.~R., {Hudson}, H.~S., {et~al.} 2011, \ssr, 159, 19

\bibitem[{{Gallagher} {et~al.}(2002){Gallagher}, {Dennis}, {Krucker},
  {Schwartz}, \& {Tolbert}}]{Gallagher2002}
{Gallagher}, P.~T., {Dennis}, B.~R., {Krucker}, S., {Schwartz}, R.~A., \&
  {Tolbert}, A.~K. 2002, \solphys, 210, 341

\bibitem[{{Gilbert} {et~al.}(2007){Gilbert}, {Alexander}, \&
  {Liu}}]{Gilbert2007}
{Gilbert}, H.~R., {Alexander}, D., \& {Liu}, R. 2007, \solphys, 245, 287

\bibitem[{{Gilbert} {et~al.}(2001){Gilbert}, {Holzer}, \&
  {Burkepile}}]{Gilbert2001}
{Gilbert}, H.~R., {Holzer}, T.~E., \& {Burkepile}, J.~T. 2001, \apj, 549, 1221

\bibitem[{{Gosain}(2012)}]{Gosain2012}
{Gosain}, S. 2012, \apj, 749, 85

\bibitem[{{Guo} {et~al.}(2012){Guo}, {Emslie}, {Massone}, \& {Piana}}]{Guo2012}
{Guo}, J., {Emslie}, A.~G., {Massone}, A.~M., \& {Piana}, M. 2012, \apj, 755,
  32

\bibitem[{{Handy} {et~al.}(1999){Handy}, {Acton}, {Kankelborg}, {Wolfson},
  {Akin}, {Bruner}, {Caravalho}, {Catura}, {Chevalier}, {Duncan}, {Edwards},
  {Feinstein}, {Freeland}, {Friedlaender}, {Hoffmann}, {Hurlburt}, {Jurcevich},
  {Katz}, {Kelly}, {Lemen}, {Levay}, {Lindgren}, {Mathur}, {Meyer}, {Morrison},
  {Morrison}, {Nightingale}, {Pope}, {Rehse}, {Schrijver}, {Shine}, {Shing},
  {Strong}, {Tarbell}, {Title}, {Torgerson}, {Golub}, {Bookbinder}, {Caldwell},
  {Cheimets}, {Davis}, {Deluca}, {McMullen}, {Warren}, {Amato}, {Fisher},
  {Maldonado}, \& {Parkinson}}]{Handy1999}
{Handy}, B.~N., {Acton}, L.~W., {Kankelborg}, C.~C., {et~al.} 1999, \solphys,
  187, 229

\bibitem[{{Hirayama}(1974)}]{Hirayama1974}
{Hirayama}, T. 1974, \solphys, 34, 323

\bibitem[{{Holman} {et~al.}(2003){Holman}, {Sui}, {Schwartz}, \&
  {Emslie}}]{Holman2003}
{Holman}, G.~D., {Sui}, L., {Schwartz}, R.~A., \& {Emslie}, A.~G. 2003, \apjl,
  595, L97

\bibitem[{{Hudson}(1991)}]{Hudson1991}
{Hudson}, H.~S. 1991, in Bulletin of the American Astronomical Society,
  Vol.~23, Bulletin of the American Astronomical Society, 1064

\bibitem[{{Hudson}(2000)}]{Hudson2000}
{Hudson}, H.~S. 2000, \apjl, 531, L75

\bibitem[{{Hurford} {et~al.}(2002){Hurford}, {Schmahl}, {Schwartz}, {Conway},
  {Aschwanden}, {Csillaghy}, {Dennis}, {Johns-Krull}, {Krucker}, {Lin},
  {McTiernan}, {Metcalf}, {Sato}, \& {Smith}}]{Hurford2002}
{Hurford}, G.~J., {Schmahl}, E.~J., {Schwartz}, R.~A., {et~al.} 2002, \solphys,
  210, 61

\bibitem[{{Ireland} {et~al.}(2013){Ireland}, {Tolbert}, {Schwartz}, {Holman},
  \& {Dennis}}]{Ireland2013}
{Ireland}, J., {Tolbert}, A.~K., {Schwartz}, R.~A., {Holman}, G.~D., \&
  {Dennis}, B.~R. 2013, \apj, 769, 89

\bibitem[{{Ji} {et~al.}(2007){Ji}, {Huang}, \& {Wang}}]{Ji2007}
{Ji}, H., {Huang}, G., \& {Wang}, H. 2007, \apj, 660, 893

\bibitem[{{Ji} {et~al.}(2003){Ji}, {Wang}, {Schmahl}, {Moon}, \&
  {Jiang}}]{Ji2003}
{Ji}, H., {Wang}, H., {Schmahl}, E.~J., {Moon}, Y.-J., \& {Jiang}, Y. 2003,
  \apjl, 595, L135

\bibitem[{{Joshi} {et~al.}(2013){Joshi}, {Kushwaha}, {Cho}, \&
  {Veronig}}]{Joshi2013}
{Joshi}, B., {Kushwaha}, U., {Cho}, K.-S., \& {Veronig}, A.~M. 2013, \apj, 771,
  1

\bibitem[{{Joshi} {et~al.}(2007){Joshi}, {Manoharan}, {Veronig}, {Pant}, \&
  {Pandey}}]{Joshi2007}
{Joshi}, B., {Manoharan}, P.~K., {Veronig}, A.~M., {Pant}, P., \& {Pandey}, K.
  2007, \solphys, 242, 143

\bibitem[{{Joshi} {et~al.}(2011){Joshi}, {Veronig}, {Lee}, {Bong}, {Tiwari}, \&
  {Cho}}]{Joshi2011}
{Joshi}, B., {Veronig}, A.~M., {Lee}, J., {et~al.} 2011, \apj, 743, 195

\bibitem[{{Joshi} {et~al.}(2009){Joshi}, {Veronig}, {Cho}, {Bong}, {Somov},
  {Moon}, {Lee}, {Manoharan}, \& {Kim}}]{Joshi2009}
{Joshi}, B., {Veronig}, A., {Cho}, K.-S., {et~al.} 2009, \apj, 706, 1438

\bibitem[{{Joshi} {et~al.}(2014){Joshi}, {Magara}, \& {Inoue}}]{JoshiNC2014}
{Joshi}, N.~C., {Magara}, T., \& {Inoue}, S. 2014, \apj, 795, 4

\bibitem[{{Karlick{\'y}}(2014)}]{Karlicky2014}
{Karlick{\'y}}, M. 2014, Research in Astronomy and Astrophysics, 14, 753

\bibitem[{{Kontar} {et~al.}(2011){Kontar}, {Brown}, {Emslie}, {Hajdas},
  {Holman}, {Hurford}, {Ka{\v s}parov{\'a}}, {Mallik}, {Massone}, {McConnell},
  {Piana}, {Prato}, {Schmahl}, \& {Suarez-Garcia}}]{Kontar2011}
{Kontar}, E.~P., {Brown}, J.~C., {Emslie}, A.~G., {et~al.} 2011, \ssr, 159, 301

\bibitem[{{Kopp} \& {Pneuman}(1976)}]{Kopp1976}
{Kopp}, R.~A., \& {Pneuman}, G.~W. 1976, \solphys, 50, 85

\bibitem[{{Kumar} {et~al.}(2012){Kumar}, {Cho}, {Bong}, {Park}, \&
  {Kim}}]{Kumar2012}
{Kumar}, P., {Cho}, K.-S., {Bong}, S.-C., {Park}, S.-H., \& {Kim}, Y.~H. 2012,
  \apj, 746, 67

\bibitem[{{Kumar} {et~al.}(2011){Kumar}, {Srivastava}, {Filippov},
  {Erd{\'e}lyi}, \& {Uddin}}]{Kumar2011}
{Kumar}, P., {Srivastava}, A.~K., {Filippov}, B., {Erd{\'e}lyi}, R., \&
  {Uddin}, W. 2011, \solphys, 272, 301

\bibitem[{{Kundu} {et~al.}(2004){Kundu}, {Garaimov}, {White}, \&
  {Krucker}}]{Kundu2004}
{Kundu}, M.~R., {Garaimov}, V.~I., {White}, S.~M., \& {Krucker}, S. 2004, \apj,
  600, 1052

\bibitem[{{Kundu} {et~al.}(2006){Kundu}, {Schmahl}, {Grigis}, {Garaimov}, \&
  {Shibasaki}}]{Kundu2006}
{Kundu}, M.~R., {Schmahl}, E.~J., {Grigis}, P.~C., {Garaimov}, V.~I., \&
  {Shibasaki}, K. 2006, \aap, 451, 691

\bibitem[{{Kushwaha} {et~al.}(2014){Kushwaha}, {Joshi}, {Cho}, {Veronig},
  {Tiwari}, \& {Mathew}}]{Kushwaha2014}
{Kushwaha}, U., {Joshi}, B., {Cho}, K.-S., {et~al.} 2014, \apj, 791, 23

\bibitem[{{Li} {et~al.}(2005){Li}, {Mickey}, \& {LaBonte}}]{LiJ2005}
{Li}, J., {Mickey}, D.~L., \& {LaBonte}, B.~J. 2005, \apj, 620, 1092

\bibitem[{{Li} \& {Gan}(2006)}]{Li2006}
{Li}, Y.~P., \& {Gan}, W.~Q. 2006, \apjl, 644, L97

\bibitem[{{Lin} {et~al.}(2002){Lin}, {Dennis}, {Hurford}, {Smith}, {Zehnder},
  {Harvey}, {Curtis}, {Pankow}, {Turin}, {Bester}, {Csillaghy}, {Lewis},
  {Madden}, {van Beek}, {Appleby}, {Raudorf}, {McTiernan}, {Ramaty}, {Schmahl},
  {Schwartz}, {Krucker}, {Abiad}, {Quinn}, {Berg}, {Hashii}, {Sterling},
  {Jackson}, {Pratt}, {Campbell}, {Malone}, {Landis}, {Barrington-Leigh},
  {Slassi-Sennou}, {Cork}, {Clark}, {Amato}, {Orwig}, {Boyle}, {Banks},
  {Shirey}, {Tolbert}, {Zarro}, {Snow}, {Thomsen}, {Henneck}, {McHedlishvili},
  {Ming}, {Fivian}, {Jordan}, {Wanner}, {Crubb}, {Preble}, {Matranga}, {Benz},
  {Hudson}, {Canfield}, {Holman}, {Crannell}, {Kosugi}, {Emslie}, {Vilmer},
  {Brown}, {Johns-Krull}, {Aschwanden}, {Metcalf}, \& {Conway}}]{Lin2002}
{Lin}, R.~P., {Dennis}, B.~R., {Hurford}, G.~J., {et~al.} 2002, \solphys, 210,
  3

\bibitem[{{Liu} {et~al.}(2009{\natexlab{a}}){Liu}, {Wang}, \&
  {Alexander}}]{Liu2009}
{Liu}, R., {Wang}, H., \& {Alexander}, D. 2009{\natexlab{a}}, \apj, 696, 121

\bibitem[{{Liu} {et~al.}(2004){Liu}, {Jiang}, {Liu}, \& {Petrosian}}]{Liu2004}
{Liu}, W., {Jiang}, Y.~W., {Liu}, S., \& {Petrosian}, V. 2004, \apjl, 611, L53

\bibitem[{{Liu} {et~al.}(2009{\natexlab{b}}){Liu}, {Petrosian}, {Dennis}, \&
  {Holman}}]{Liu2009a}
{Liu}, W., {Petrosian}, V., {Dennis}, B.~R., \& {Holman}, G.~D.
  2009{\natexlab{b}}, \apj, 693, 847

\bibitem[{{Metcalf} {et~al.}(1996){Metcalf}, {Hudson}, {Kosugi}, {Puetter}, \&
  {Pina}}]{Metcalf1996}
{Metcalf}, T.~R., {Hudson}, H.~S., {Kosugi}, T., {Puetter}, R.~C., \& {Pina},
  R.~K. 1996, \apj, 466, 585

\bibitem[{{Moore} \& {Roumeliotis}(1992)}]{Moore1992}
{Moore}, R.~L., \& {Roumeliotis}, G. 1992, in Lecture Notes in Physics, Berlin
  Springer Verlag, Vol. 399, IAU Colloq. 133: Eruptive Solar Flares, ed.
  Z.~{Svestka}, B.~V. {Jackson}, \& M.~E. {Machado}, 69

\bibitem[{{Mrozek}(2011)}]{Mrozek2011}
{Mrozek}, T. 2011, \solphys, 270, 191

\bibitem[{{Nakajima} {et~al.}(1994){Nakajima}, {Nishio}, {Enome}, {Shibasaki},
  {Takano}, {Hanaoka}, {Torii}, {Sekiguchi}, {Bushimata}, {Kawashima},
  {Shinohara}, {Irimajiri}, {Koshiishi}, {Kosugi}, {Shiomi}, {Sawa}, \&
  {Kai}}]{Nakajima1994}
{Nakajima}, H., {Nishio}, M., {Enome}, S., {et~al.} 1994, IEEE Proceedings, 82,
  705

\bibitem[{{Netzel} {et~al.}(2012){Netzel}, {Mrozek}, {Ko{\l}oma{\'n}ski}, \&
  {Gburek}}]{Netzel2012}
{Netzel}, A., {Mrozek}, T., {Ko{\l}oma{\'n}ski}, S., \& {Gburek}, S. 2012,
  \aap, 548, A89

\bibitem[{{Patsourakos} {et~al.}(2013){Patsourakos}, {Vourlidas}, \&
  {Stenborg}}]{Patsourakos2013}
{Patsourakos}, S., {Vourlidas}, A., \& {Stenborg}, G. 2013, \apj, 764, 125

\bibitem[{{Priest} \& {Forbes}(2002)}]{Priest2002}
{Priest}, E.~R., \& {Forbes}, T.~G. 2002, \aapr, 10, 313

\bibitem[{{Rust}(2003)}]{Rust2003}
{Rust}, D.~M. 2003, Advances in Space Research, 32, 1895

\bibitem[{{Saint-Hilaire} \& {Benz}(2005)}]{Saint-Hilaire2005}
{Saint-Hilaire}, P., \& {Benz}, A.~O. 2005, \aap, 435, 743

\bibitem[{{Scherrer} {et~al.}(1995){Scherrer}, {Bogart}, {Bush}, {Hoeksema},
  {Kosovichev}, {Schou}, {Rosenberg}, {Springer}, {Tarbell}, {Title},
  {Wolfson}, {Zayer}, \& {MDI Engineering Team}}]{Scherrer1995}
{Scherrer}, P.~H., {Bogart}, R.~S., {Bush}, R.~I., {et~al.} 1995, \solphys,
  162, 129

\bibitem[{{Shibata}(1998)}]{Shibata1998}
{Shibata}, K. 1998, \apss, 264, 129

\bibitem[{{Sim{\~o}es} {et~al.}(2013){Sim{\~o}es}, {Fletcher}, {Hudson}, \&
  {Russell}}]{Simoes2013}
{Sim{\~o}es}, P.~J.~A., {Fletcher}, L., {Hudson}, H.~S., \& {Russell}, A.~J.~B.
  2013, \apj, 777, 152

\bibitem[{{Smith} {et~al.}(2002){Smith}, {Lin}, {Turin}, {Curtis}, {Primbsch},
  {Campbell}, {Abiad}, {Schroeder}, {Cork}, {Hull}, {Landis}, {Madden},
  {Malone}, {Pehl}, {Raudorf}, {Sangsingkeow}, {Boyle}, {Banks}, {Shirey}, \&
  {Schwartz}}]{Smith2002}
{Smith}, D.~M., {Lin}, R.~P., {Turin}, P., {et~al.} 2002, \solphys, 210, 33

\bibitem[{{Sturrock}(1966)}]{Sturrock1966}
{Sturrock}, P.~A. 1966, \nat, 211, 695

\bibitem[{{Sui} \& {Holman}(2003)}]{Sui2003}
{Sui}, L., \& {Holman}, G.~D. 2003, \apjl, 596, L251

\bibitem[{{Sui} {et~al.}(2004){Sui}, {Holman}, \& {Dennis}}]{Sui2004}
{Sui}, L., {Holman}, G.~D., \& {Dennis}, B.~R. 2004, \apj, 612, 546

\bibitem[{{Sui} {et~al.}(2005){Sui}, {Holman}, {White}, \& {Zhang}}]{Sui2005}
{Sui}, L., {Holman}, G.~D., {White}, S.~M., \& {Zhang}, J. 2005, \apj, 633,
  1175

\bibitem[{{Takano} {et~al.}(1997){Takano}, {Nakajima}, {Enome}, {Shibasaki},
  {Nishio}, {Hanaoka}, {Shiomi}, {Sekiguchi}, {Kawashima}, {Bushimata},
  {Shinohara}, {Torii}, {Fujiki}, \& {Irimajiri}}]{Takano1997}
{Takano}, T., {Nakajima}, H., {Enome}, S., {et~al.} 1997, in Lecture Notes in
  Physics, Berlin Springer Verlag, Vol. 483, Coronal Physics from Radio and
  Space Observations, ed. G.~{Trottet}, 183

\bibitem[{{T{\"o}r{\"o}k} \& {Kliem}(2005)}]{Torok2005}
{T{\"o}r{\"o}k}, T., \& {Kliem}, B. 2005, \apjl, 630, L97

\bibitem[{{Vemareddy} \& {Zhang}(2014)}]{Vema2014}
{Vemareddy}, P., \& {Zhang}, J. 2014, ArXiv e-prints, arXiv:1410.2158

\bibitem[{{Veronig} {et~al.}(2002){Veronig}, {Vr{\v s}nak}, {Temmer}, \&
  {Hanslmeier}}]{Veronig2002}
{Veronig}, A., {Vr{\v s}nak}, B., {Temmer}, M., \& {Hanslmeier}, A. 2002,
  \solphys, 208, 297

\bibitem[{{Veronig} {et~al.}(2005){Veronig}, {Brown}, {Dennis}, {Schwartz},
  {Sui}, \& {Tolbert}}]{Veronig2005}
{Veronig}, A.~M., {Brown}, J.~C., {Dennis}, B.~R., {et~al.} 2005, \apj, 621,
  482

\bibitem[{{Veronig} {et~al.}(2006){Veronig}, {Karlick{\'y}}, {Vr{\v s}nak},
  {Temmer}, {Magdaleni{\'c}}, {Dennis}, {Otruba}, \& {P{\"o}tzi}}]{Veronig2006}
{Veronig}, A.~M., {Karlick{\'y}}, M., {Vr{\v s}nak}, B., {et~al.} 2006, \aap,
  446, 675

\bibitem[{{Warmuth} \& {Mann}(2013)}]{Warmuth2013}
{Warmuth}, A., \& {Mann}, G. 2013, \aap, 552, A87

\bibitem[{{White} {et~al.}(2011){White}, {Benz}, {Christe},
  {F{\'a}rn{\'{\i}}k}, {Kundu}, {Mann}, {Ning}, {Raulin}, {Silva-V{\'a}lio},
  {Saint-Hilaire}, {Vilmer}, \& {Warmuth}}]{White2011}
{White}, S.~M., {Benz}, A.~O., {Christe}, S., {et~al.} 2011, \ssr, 159, 225

\bibitem[{{Zhang} {et~al.}(2014){Zhang}, {Tan}, {Karlick{\'y}},
  {M{\'e}sz{\'a}rosov{\'a}}, {Huang}, {Tan}, \& {Sim{\~o}es}}]{Zhang2014}
{Zhang}, Y., {Tan}, B., {Karlick{\'y}}, M., {et~al.} 2014, ArXiv e-prints,
  arXiv:1411.4766

\end{thebibliography}

\clearpage

\end{document}